\documentclass[12pt]{article}

\pdfoutput=1

\usepackage[top=80pt,bottom=85pt,left=85pt,right=85pt]{geometry}
\usepackage{amsmath,amssymb,graphicx,float,color,tikz,subfigure,cite}
\usepackage[debug,pageanchor=false]{hyperref}
\definecolor{link}{rgb}{.8,.15,.1}
\hypersetup{colorlinks=true,linkcolor=link,citecolor=link,urlcolor=link,linktocpage}

\usepackage[vcentermath]{youngtab}

\makeatletter
\@addtoreset{equation}{section}
\makeatother

\begin{document}

	\begin{titlepage}
	
	\begin{flushright}
        {CERN-TH-2018-229}\\
        \end{flushright}

	\begin{center}

	\vskip .5in 
	\noindent

	{\Large \bf{Holographic duals of 6d RG flows}}

	\bigskip\medskip

	G.~Bruno De Luca,$^{1}$ Alessandra Gnecchi,$^2$\\ Gabriele Lo Monaco,$^{1}$ Alessandro Tomasiello$^{1}$\\

	\bigskip\medskip
	{\small 

	$^1$ Dipartimento di Fisica, Universit\`a di Milano--Bicocca, \\ Piazza della Scienza 3, I-20126 Milano, Italy \\ and \\ INFN, sezione di Milano--Bicocca
	\\	
	\vspace{.3cm}
	$^2$ Theoretical Physics Department, CERN, Geneva, Switzerland

	}

	\vskip .5cm 
	{\small \tt g.deluca8@campus.unimib.it, alessandra.gnecchi@cern.ch, g.lomonaco1@campus.unimib.it, alessandro.tomasiello@unimib.it}
	\vskip .9cm 
	     	{\bf Abstract }
	
	\vskip .1in
	\end{center}

	\noindent

	A notable class of superconformal theories (SCFTs) in six dimensions is parameterized by an integer $N$, an ADE group $G$, and two nilpotent elements $\mu_\mathrm{L,R}$ in $G$. Nilpotent elements have a natural partial ordering, which has been conjectured to coincide with the hierarchy of renormalization-group flows among the SCFTs. In this paper we test this conjecture for $G=\mathrm{SU}(k)$, where AdS$_7$ duals exist in IIA. We work with a seven-dimensional gauged supergravity, consisting of the gravity multiplet and two $\mathrm{SU}(k)$ non-Abelian vector multiplets. We show that this theory has many supersymmetric AdS$_7$ vacua, determined by two nilpotent elements, which are naturally interpreted as IIA AdS$_7$ solutions. The BPS equations for domain walls connecting two such vacua can be solved analytically, up to a Nahm equation with certain boundary conditions. The latter admit a solution connecting two vacua if and only if the corresponding nilpotent elements are related by the natural partial ordering, in agreement with the field theory conjecture.

	\noindent

	\vfill
	\eject

	\end{titlepage}

\tableofcontents

\section{Introduction} 
\label{sec:intro}

There is by now a lot of evidence for the existence of a class of six-dimensional SCFTs ${\cal T}^N_{G,\mu_{\rm L}, \mu_{\rm R}}$, characterized by an integer $N$, an ADE group $G$ and two nilpotent elements $\mu_{{\rm L,R}}$ in $G$. 
For $G={\rm SU}(k)$, these theories were proposed long ago \cite{intriligator-6d,intriligator-6d-II,brunner-karch,hanany-zaffaroni-6d}; their holographic duals were found in \cite{afrt,10letter,cremonesi-t} with growing amount of detail. For $G={\rm SO}(2k)$ or $E_k$, the theories were suggested to exist in \cite{dhtv} and found in \cite{heckman-rudelius-t}. In \cite{heckman-rudelius-t2} it was found that a certain generalization of ${\cal T}^N_{G,\mu_{\rm L}, \mu_{\rm R}}$ involving two (non necessarily ADE) groups in fact covers the space of all possible SCFTs with a large enough number of tensor multiplets.

It was also proposed in \cite{gaiotto-t-6d,heckman-rudelius-t,heckman-rudelius-t2} that two theories are connected by a renormalization group (RG) flow if and only if the corresponding nilpotent elements are related by partial ordering. Focusing on the left nilpotent element:
\begin{equation}\label{eq:RG}
	{\cal T}^N_{G,\mu_{\rm L},\mu_{\rm R}} \buildrel{\text{RG}}\over\longrightarrow {\cal T}^N_{G,\mu'_{\rm L},\mu_{\rm R}} \qquad \Leftrightarrow \qquad \mu_{\rm L} < \mu'_{\rm L}\,,
\end{equation}
where on the right hand side $<$ represents the natural partial ordering among nilpotent elements, to be reviewed below. 

In this paper, we will test this conjecture for $G= {\rm SU}(k)$ using supergravity, by finding BPS solutions that interpolate between two AdS$_7$ vacua. While those were found directly in IIA supergravity, we will work with an effective seven-dimensional description that contains all the expected vacua with a given $k$. 

It was already found in \cite{prt} that for every AdS$_7$ solution there is a consistent truncation to the so-called minimal gauged supergravity in seven dimensions.\footnote{\cite{malek-samtleben-vallcamell} recently showed how to reproduce (or explain) that consistent truncation from the point of view of exceptional field theory.} The fact that this theory is the same for all AdS$_7$ vacua seems to indicate that it captures some kind of universal sector common to all of them and to their CFT$_6$ duals. While this theory is interesting and useful, it cannot be used to describe RG flows between two different theories, since in this description all the vacua are identified with one another. In order to tell them apart, we would need a reduction where more modes of the internal space are kept. In particular, as already suggested in \cite{prt}, one might want to include in the reduction the modes living on the D6s and D8s present in the AdS$_7$ solution. 

We will not work out this reduction, but it is easy to guess what the result would be. Let us start from the theory which is at the top of the RG hierarchy. For $G={\rm SU}(k)$, nilpotent elements are associated to Young diagrams, as we will see in detail in section \ref{sec:6d}. The theory at the top of the RG chain is obtained by taking $\mu_{\rm L}$ and $\mu_{\rm R}$ to be a vertical stack of $k$ boxes (for example, for $k=4$, $\mu_{\rm L}=\mu_{\rm R}=${\tiny $\yng(1,1,1,1)$}). The dual of this theory has two stacks of $k$ D6-branes. (The internal manifold $M_3$ has the topology of $S^3$, and the two stacks sit at the north and south poles respectively.) This suggests the presence of two ${\rm SU}(k)$ vector multiplets in the seven-dimensional effective theory. Indeed the SCFT in this case has ${\rm SU}(k)\times {\rm SU}(k)$ flavor symmetry. 

For other AdS$_7$ solutions, the number $k$ can still be identified as a certain flux quantum of the RR field $F_2$, but the gauge groups of the effective seven-dimensional supergravity (or the flavor symmetry of the dual CFT) is a subgroup of ${\rm SU}(k)\times {\rm SU}(k)$. We will argue that these solutions are represented in the theory with ${\rm SU}(k)\times {\rm SU}(k)$ vector multiplets by vacua where the gauge symmetry has been partially Higgsed. Indeed we will see that seven-dimensional minimal gauged supergravity coupled to two ${\rm SU}(k)$ vector multiplets has many AdS$_7$ vacua, each associated to a choice of two Young diagrams with $k$ boxes.\footnote{\label{foot:prev}One particular case, for $k=2$, had already been found in \cite{Karndumri:2014hma}, along with an RG flow connecting it to the trivial vacuum; this was in fact one of the inspirations of this paper. The vacua of minimal supergravity coupled to extra vectors was also considered in a related context by \cite{danielsson-dibitetto-vargas-swamp}, where it was concluded that non-supersymmetric vacua have tachyons.} They are non-abelian vacua, in the sense that the scalars form a reducible $\mathrm{SU}(2)$ representation. It is very natural to surmise that these are exactly the AdS$_7$ solutions of \cite{afrt,10letter,cremonesi-t}, for a fixed $k$. The $\mathrm{SU}(2)$ representation is interpreted as a ``puffing up'' process whereby the D6-branes become D8s, in a Myers-like \cite{myers} process.

Having found seven-dimensional avatars of all the AdS$_7$ vacua for a given $k$, we then proceed to look for BPS domain-wall solutions that connect them. According to the rules of holography, these should be dual to the RG flows (\ref{eq:RG}). We generalize the vacuum Ansatz to let the scalars and geometry change with the radial coordinate of AdS$_7$. 

With this relaxed Ansatz, the BPS equations of our seven-dimensional gauged supergravity reduce to a variant of Nahm's equations \cite{nahm-eq}. They were indeed expected to play a role in the study of the ${\cal T}^N_{G,\mu_{\rm L}, \mu_{\rm R}}$ theories, for reasons similar to their appearance in the study of 3d	 theories \cite{gaiotto-witten-1}. Using results in \cite{kronheimer-nahmeq}, we conclude that a BPS domain wall exists exactly when predicted by the field theory conjecture (\ref{eq:RG}), thus strongly validating it. 
 
While domain walls connecting two different AdS vacua are by now routinely found in several dimensions, our result is notable in that we have a large number of vacua and an even larger of domain walls --- both in fact growing arbitrarily large with $k$. Most of the times in the literature the BPS equations have to be solved numerically. In our case, we are able to make contact with the well-studied Nahm equations, and that allows us to both avoid a numerical study and prove the existence of a large number of domain-wall solutions.
 
Our seven-dimensional supergravity approach was enough to capture most of the relevant field theory physics; we should stress, however, that we have not found a consistent truncation relating it to IIA supergravity in ten dimensions. That would allow us to uplift the domain walls to ten dimensions. One obstacle to find such a consistent truncation has to do with higher-derivative terms. In ten dimensions, the IIA supergravity action where the AdS$_7$ solutions \cite{afrt,10letter,cremonesi-t} were found only contains two derivatives; however, the brane (open-string) action to which it couples contains a Dirac--Born--Infeld (DBI) and a Chern--Simons term. This is where the $\mathrm{SU}(k)$ vector field lives, and it appears with higher derivatives. In fact the non-abelian DBI that we would need has not even been worked out in full. On the other hand, in our seven-dimensional theory the non-abelian $\mathrm{SU}(k)$ vectors appear with two derivatives only, as is customary in a gauged supergravity. To find a consistent truncation to IIA, one should extend the action for the vectors to a DBI-like supersymmetric action, perhaps along the lines of \cite{andrianopoli-dauria-trigiante-DBI}. 

The paper is organized as follows. In section \ref{sec:6d} we give a lightning review of the six-dimensional SCFTs we are interested in, and of some features of nilpotent elements in ADE groups. In section \ref{sec:7dsug} we describe the seven-dimensional supergravity action we will use. In section \ref{sec:vacua} we will find BPS AdS$_7$ vacua for this theory. For simplicity and clarity at this stage we will find vacua where only one of the two $\mathrm{SU}(k)$ is spontaneously broken, while the other is untouched; this corresponds to keeping one of the two Young diagrams (say $\mu_\mathrm{R}$) trivial, while varying the other. Thus our vacua in this section are determined by the choice of only one Young diagram $\mu_\mathrm{L}$.
We will then look for domain walls among these vacua in section \ref{sec:dw}, still keeping one of the two $\mathrm{SU}(k)$ untouched. Finally in section \ref{sec:double} we will generalize the results of the previous two sections to the most general case where both $\mathrm{SU}(k)$ gauge groups are broken; here both $\mu_\mathrm{L}$ and $\mu_\mathrm{R}$ will be nontrivial.


\section{The field theories} 
\label{sec:6d}

We will begin with a quick review of the six-dimensional SCFTs that we are going to investigate holographically. A longer discussion of the field theories and their AdS$_7$ duals can be found in \cite[Sec.~2]{cremonesi-t}. 

The SCFTs 
\begin{equation}\label{eq:T}
	{\cal T}^N_{G,\mu_{\rm L},\mu_{\rm R}}
\end{equation}
are associated to a positive integer $N$, an ADE Lie group $G$, and two nilpotent elements $\mu_\mathrm{L}$, $\mu_\mathrm{R}\in G$. The case of interest to this paper will be $G=\mathrm{SU}(k)$.

When the nilpotent elements are zero, we have the theory ${\cal T}^N_{G,0,0}$. This has ${\mathcal N}=(1,0)$ supersymmetry, and a flavor symmetry $G_\mathrm{L}\times G_\mathrm{R}$ consisting of two copies of $G$ and of one $\mathrm{U}(1)$. (The $\mathrm{U}(1)$ will play no role in what follows, and we will ignore it.) This SCFT is engineered in M-theory by $N+1$ M5-branes on a $\mathbb{C}^2/\Gamma_G$ singularity, where $\Gamma_G$ is the discrete subgroup of $\mathrm{SU}(2)$ associated to $G$ by the McKay correspondence. For example, for $G=\mathrm{SU}(k)$ of interest in this paper, $\Gamma_G= \mathbb{Z}_k$. Another possible realization is in IIA, by considering $N+1$ NS5-branes on $k$ D6-branes \cite{hanany-zaffaroni,brunner-karch}, or in IIB with $k$ D5-branes and a $\mathbb{C}^2/\mathbb{Z}_{N+1}$ singularity \cite{intriligator-6d}. 

The more general SCFTs (\ref{eq:T}) with $\mu_\mathrm{L,R}\neq 0$ has still ${\mathcal N}=(1,0)$ supersymmetry, but its flavor symmetry is now broken to the commutant of $\mu_\mathrm{L}$ in $G_\mathrm{L}$, times the commutant of $\mu_\mathrm{R}$ inside $G_R$. Two nilpotent elements which can be brought to one another by the adjoint action of $G$ produce the same theory: ${\cal T}^N_{G,\mu_{\rm L},\mu_{\rm R}}\equiv {\cal T}^N_{G,\mu_{\rm L}',\mu_{\rm R}} $ $\Leftrightarrow \exists g\in G | g \mu_\mathrm{L} g^{-1}= \mu_\mathrm{L}'$. Two such nilpotent elements are said to belong to the same \emph{nilpotent orbit} ${\mathcal O}$ of $G$. So what really matters in (\ref{eq:T}) is not the $\mu_\mathrm{L,R}$ themselves, but the nilpotent orbits ${\mathcal O}_\mathrm{L,R}$ to which they belong, or of which they are a representative.

These more general SCFTs can also be engineered in string theory. For $G=\mathrm{SU}(k)$, (\ref{eq:T})  can be engineered in IIA by adding D8-branes on which the D6s end. Some choices of $\mu_\mathrm{L,R}$ for $G=\mathrm{SO}(2k)$ can also be realized by adding O6-planes \cite{hanany-zaffaroni}. For the remaining choices of $\mu_\mathrm{L,R}$, and for all the $G=E_6$, $E_7$, $E_8$ cases, there is an engineering in F-theory, as predicted in \cite{dhtv} and realized in \cite{heckman-rudelius-t,mekareeya-rudelius-t}.

The string realization suggests that the theories with a given $N$ and $G$ are related by Higgs RG flows \cite{gaiotto-t-6d,heckman-rudelius-t}. According to this conjecture, each theory (\ref{eq:T}) can be viewed as the result of having partially Higgsed the flavor symmetry $G_\mathrm{L}\times G_\mathrm{R}$ of ${\cal T}^N_{G,0,0}$. The Higgs moduli space of ${\cal T}^N_{G,0,0}$ has quaternionic dimension \cite{mekareeya-ohmori-shimizu-t}
\begin{equation}
	N+1 + \mathrm{dim}(G)\,.
\end{equation}
The structure of this moduli space is not completely known, but the $\mathrm{dim}(G)$ directions are supposed to be related to the space ${\mathcal N}$ of nilpotent elements in $G$ (also known as the \emph{nilpotent cone}). This space has many singularities; if one switches on a vacuum expectation value (vev) corresponding to the points in moduli space on such a singularity, and one follows the RG flow, one expects to obtain a new SCFT in the infrared. (Choosing a smooth point is expected to lead to a free theory in the infrared.) 

Given a point $\mu\in{\mathcal N}$, the type of singularity depends on its orbit ${\mathcal O}$.  Choosing an orbit ${\mathcal O}_\mathrm{L,R}$ for both factors of $G$ of the flavor symmetry group $G_\mathrm{L}\times G_\mathrm{R}$ of ${\cal T}^N_{G,0,0}$ results then in the general theory (\ref{eq:T}). The Higgs moduli space dimension is now reduced to 
\begin{equation}\label{eq:dimH-red}
	N+1 + \mathrm{dim}(G)- \mathrm{dim}({\mathcal O}_\mathrm{L}) - \mathrm{dim}({\mathcal O}_\mathrm{R})\,.
\end{equation}

Even the reduced moduli space of dimension (\ref{eq:dimH-red}) will have singularities, inherited from the original nilpotent cone ${\mathcal N}$. So it will be possible to choose again a vev corresponding to a singularity; flowing to the infrared will produce a new theory. This gives rise to a ``hierarchy'' or RG flows. 

\begin{figure}[ht]
	\centering
		\includegraphics[width=10cm]{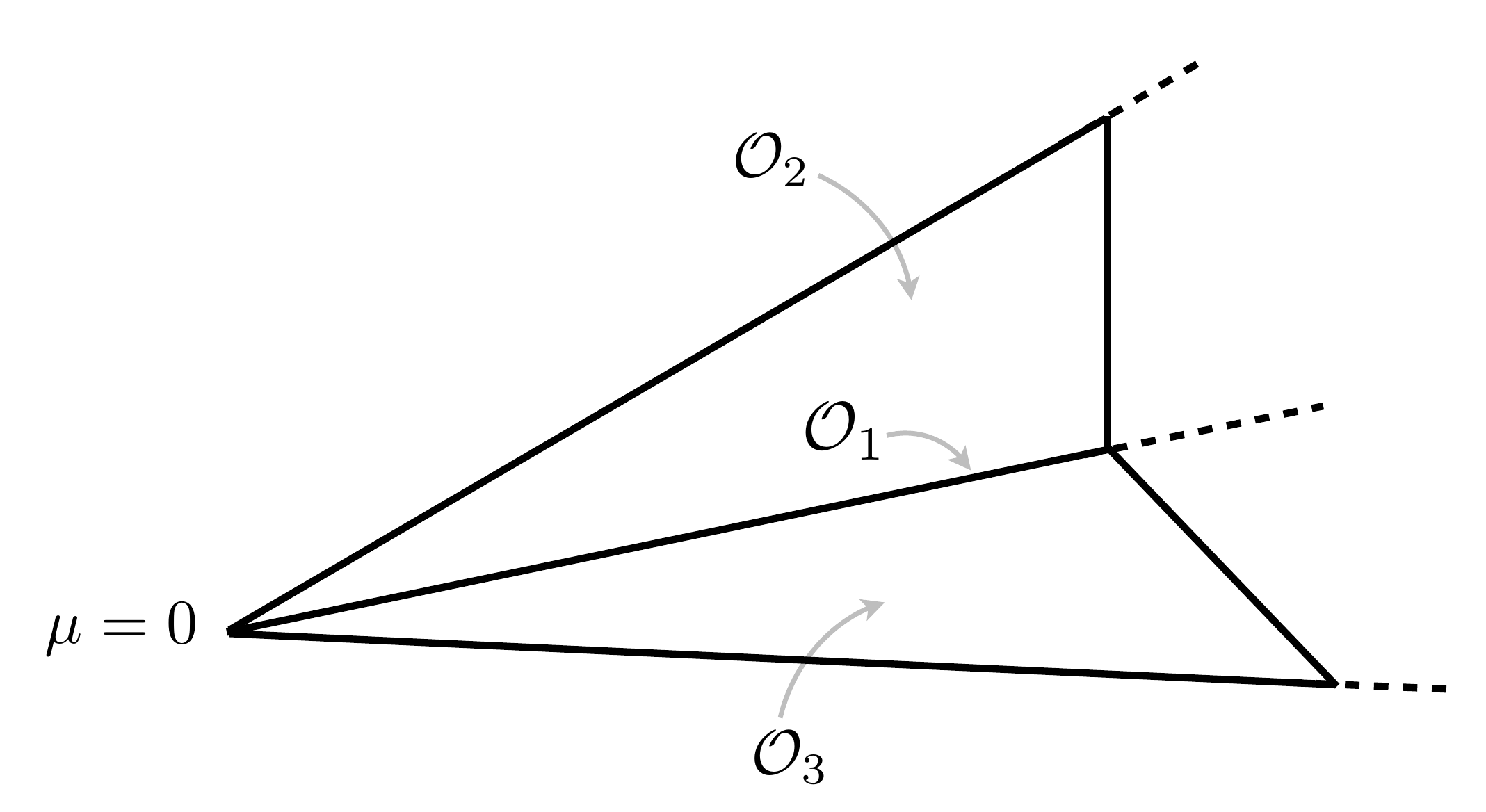}
	\caption{A sketch of the structure of the nilpotent cone. ${\mathcal O}_1$ is not meant to be included in ${\mathcal O}_2$ and ${\mathcal O}_3$, but rather in their closure.}
	\label{fig:orbits}
\end{figure}

To understand this hierarchy better, notice that there is a natural partial ordering among nilpotent orbits. An orbit ${\mathcal O}$ is \emph{larger than} (or \emph{dominates}) an orbit ${\mathcal O}'$ if ${\mathcal O}'$ belongs to the closure of ${\mathcal O}$; see Fig.~\ref{fig:orbits} for a sketch. The hierarchy of SCFTs can be now thought of as follows. One starts from the theory ${\mathcal T}^N_{G,0,0}$, where both $\mu_\mathrm{L,R}$ are the trivial nilpotent orbit $\mu=0$ (the origin of the cone in Fig.~\ref{fig:orbits}). One can Higgs the theory by choosing a vev $\mu_{L,1} \in {\mathcal O}_1$ in the moduli space ${\mathcal N}$; in the moduli space of the resulting theory ${\mathcal T}^N_{G,\mu_\mathrm{L,1},0}$, the cone ${\mathcal N}$ is now reduced to a slice which meets ${\mathcal O}_1$ in $\mu_1$. Referring again to Fig.~\ref{fig:orbits}, one can now Higgs the theory further by choosing $\mu_2 \in {\mathcal O}_2$ or by $\mu_3\in {\mathcal O}_3$, and so on.

This is the reason the arrows in (\ref{eq:RG}) go backwards: intuitively, one loses more degrees of freedom in the infrared by choosing a vev in a more generic point in the Higgs moduli space, corresponding to a larger orbit.

Let us now be more concrete and describe the nilpotent orbits for $G=\mathrm{SU}(k)$, the case of interest in this paper. (Nilpotent orbits in the D and E case have a more complicated classification \cite{collingwood-mcgovern}.) Every nilpotent element is conjugated to one of the following form:
\begin{equation}\label{eq:jordan}
	\mu= \left(\begin{array}{ccc}
		J_{d_1} & &  \\
		 & J_{d_2} &  \\
		 & & \ddots  
	\end{array}\right)\,,
	\qquad J_d\equiv 
	\left.\left(
	\begin{matrix}
	0&1&&\\
	&0&1&\\
	&&\ddots&\ddots
	\end{matrix}
	\right)\right\} d\,.
\end{equation} 
Two $\mu$ whose $d_a$ are related by permutations are in fact also conjugated; so to avoid repetitions we assume that the $d_a$ are listed in increasing order, $d_a \le d_{a+1}$. So each nilpotent orbit is identified by a \emph{partition} of $k$, namely a choice $[d_1,\, d_2,\, \ldots]$ such that $\sum_a d_a=k$. For example, the partition $[1,\,1,\,\ldots,\,1]\equiv [1^k]$ is associated to $\mu=0$ (which is indeed nilpotent), while the partition $[k]$ is associated to the single Jordan block $J_k$.

It is also common to denote these partitions by Young diagrams. One can associate the $d_a$ of the partition to either the rows or the columns of a Young diagram; both possibilities are used in the literature in different contexts (for reasons that will soon become clear). Here we are going to follow the convention that the $d_a$ are the number of boxes in each row of the Young diagram. So for example, say for $k=6$: 
\begin{equation}\label{eq:16-6}
	[1^6] \mapsto {\small\yng(1,1,1,1,1,1)} \, ,\qquad [6] \mapsto {\small\yng(6)}\,.
\end{equation}

The Young diagram representation of a partition is useful for various reasons; one is that it allows to introduce the \emph{transpose} partition $\mu^t$, which is simply obtained by reflecting it along a diagonal axis. For example with the help of (\ref{eq:16-6}) we see immediately that $[1^6]^t= [6]$, and $[6]^t=[1^6]$. Another way of defining $\mu^t$ is by counting the number of boxes in each column of the Young diagrams associated to $\mu$. The quaternonic dimension of the orbit ${\mathcal O}_\mu$ is
\begin{equation}\label{eq:dimO}
	\mathrm{dim}({\mathcal O}_\mu)= \frac12 \left(k^2-\sum_a (\mu^t_a)^2\right)\,.
\end{equation}
For example $\mathrm{dim}({\mathcal O}_{[1^k]})= \frac12(k^2-k^2)=0$, and indeed $[1^k]$ is associated to the nilpotent element $\mu=0$; while $\mathrm{dim}({\mathcal O}_{[k]})= \frac12{k^2-k}$.

The flavor symmetry of (\ref{eq:T}) now can be described combinatorially. Define
\begin{equation}\label{eq:fa}
	f^\mathrm{L,R}_a \equiv (\mu_\mathrm{L,R})^t_a-(\mu_\mathrm{L,R})^t_{a+1}\,.
\end{equation}
Notice that
\begin{equation}\label{eq:fa-blocks}
	f_a^\mathrm{L,R} = \# \{ \text{blocks } J_a \text{ with dimension } a\}\,.
\end{equation}
So for example 
{\small\begin{equation}\label{eq:1225}
	\yng(1,2,2,5)
\end{equation}}
has $d_1=1$, $d_2=2$, $d_3=2$, $d_4=5$, and $f_1=1$, $f_2=2$, $f_5=1$. Since the total dimension of $\mu$ is $k$, from (\ref{eq:fa-blocks}) we have
\begin{equation}\label{eq:afak}
	\sum_a a f_a^\mathrm{L}=  \sum_a a f_a^\mathrm{R}= k\,.
\end{equation}
 
In terms of (\ref{eq:fa}), the flavor symmetry of (\ref{eq:T}) is\footnote{The effective theory on the tensor branch might suggest a larger number of abelian factors, but many of them are anomalous; compactifications to lower dimensions also suggest a reduced number at the conformal point. (\ref{eq:flavor-flav}) is suggested naturally by the gravity duals.}
\begin{equation}\label{eq:flavor-flav}
	\mathrm{S}\left(\Pi_a \mathrm{U}(f_a^\mathrm{L})\right) \times \mathrm{S}\left(\Pi_a \mathrm{U}(f_a^\mathrm{R})\right)\,. 
\end{equation}
So for example the $\mu^t_a$ for $[1^6]\cong ${\tiny $\yng(1,1,1,1,1,1)$} are $[6]$; in this case the only non-zero $f_a$ is $f_1=6$. Indeed $\mu=0$ corresponds to the partition $[1^6]$, and the flavor symmetry for ${\mathcal T}^N_{\mathrm{SU}(6),0,0}$ is $\mathrm{SU}(6)\times \mathrm{SU}(6)$. On the other hand, the $\mu^t_a$ for $[6]\cong ${\tiny $\yng(6)$} are $[1,1,1,1,1,1]\equiv [1^6]$; in this case the only non-zero $f_a$ is $f_6=1$. So if we take this as $\mu_\mathrm{L}$, we have that the flavor symmetry of $\mathrm{T}^N_{\mathrm{SU}(6),[6],0}$ is just one $\mathrm{SU}(6)$.
 
\begin{figure}[ht]
	\centering
		\includegraphics[width=14cm]{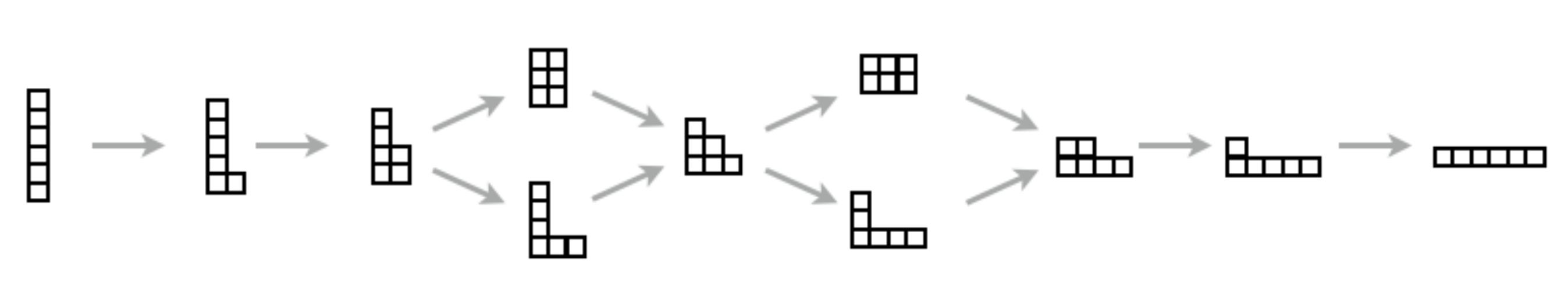}
	\caption{The hierarchy of Young diagrams with six boxes. The arrows here represent possible RG flows.}
	\label{fig:hasse-su6}
\end{figure}

The ordering of these orbits is also easy to describe. A diagram $\mu$ dominates $\mu'$ (which we often denote by $\mu > \mu'$) if $\mu'$ can be obtained from $\mu$ by removing a box from a higher row and adding it to a lower row. This is more easily described by an example: in Fig.~\ref{fig:hasse-su6} we have depicted the partial ordering among Young diagrams with $N=6$ boxes. The arrows depict possible RG flows, and thus point from smaller to larger Young diagrams. Indeed we see that on the left we have the vertical Young diagram {\tiny $\yng(1,1,1,1,1,1)$}, corresponding to the partition $[1^6]$ and thus to $\mu=0$, which belongs to the smallest possible orbit; this is depicted in the sketch of Fig.~\ref{fig:orbits} as the tip of the cone. At the right extremum of Fig.~\ref{fig:hasse-su6} we instead have the horizontal Young diagram {\tiny $\yng(6)$}, corresponding to the largest possible orbit $\mu= J_6$.

While it is customary to label the theories (\ref{eq:T}) by nilpotent elements, there is another point of view, that will be even more important for us. By the Jacobson--Morozov theorem \cite[Chap.~3]{collingwood-mcgovern}, to a nilpotent element $\mu \in G$ one can add two more elements in $G$ which together with $\mu$ satisfy the $\mathrm{sl}(2,\mathbb{C})$ commutation relations. (One can think of $\mu$ as the ``creator operator'' in such a triple.) One can then find a change of basis that takes this triple to three  Hermitian matrices $\sigma^i$ such that
\begin{equation}\label{eq:su2}
	[\sigma^i,\sigma^j] = \epsilon^{ijk} \sigma^k\,.
\end{equation}
In other words, one can associate to $\mu$ an embedding 
\begin{equation}
	\sigma:{\rm su}(2)\subset \mathfrak{g}\,,
\end{equation}
where $\mathfrak{g}$ is the Lie algebra of $G$.

Another way of thinking about the $\sigma_i$ is that they provide a reducible su(2) representation: \begin{equation}\label{eq:red-su2}
	\sigma^i=\left(\begin{array}{ccc}
		\sigma^i_{1} & &  \\
		 & \sigma^i_{2} &  \\
		 & & \ddots 
	\end{array}\right)
\end{equation}
where $\sigma^i_a$ has dimension $d_a$; in other words, a direct sum of irreducible representations of spins $\ell_1$, $\ell_2$, $\ldots$ such that $2 \ell_a + 1= d_a$ in (\ref{eq:jordan}). The blocks obey
\begin{equation}\label{eq:su2-rep}
	\mathrm{Tr}(\sigma^i_a \sigma^j_a)=-\kappa_a^2 \delta^{ij} \, ,\qquad  \kappa_a^2\equiv \frac{\ell_a(\ell_a+1)(2\ell_a+1)}{3}= \frac{d_a(d_a^2-1)}{12}\,.
\end{equation}

There are several other important aspects of the theories (\ref{eq:T}). Let us mention here for example that, while none of them has a Lagrangian description so far, many of them have an effective description in terms of gauge theories. Besides the Higgs moduli space we dealt with so far, there is a ``tensor moduli space'' (similar to the Coulomb branch for ${\mathcal N}=2$ in four dimensions); as the name suggests, giving a vev along this space results in a theory with several abelian two-index antisymmetric tensors. These tensors are coupled to several non-abelian gauge fields and hypermultiplets. The resulting theory is not renormalizable; its ultraviolet completion is the original SCFT. For $G=\mathrm{SU}(k)$, there is an easy algorithm to read off the tensor-branch effective theory we just sketched from the two Young diagrams $\mu_\mathrm{L,R}$; it is illustrated for example in \cite[Fig.~2]{cremonesi-t}. Again we refer to that reference for further details.

Finally let us mention very briefly the AdS$_7$ duals. These exist for $G=\mathrm{SU}(k)$ and with some caveats for $G=\mathrm{SO}(2k)$. They were first found numerically in \cite{afrt}, then analytically in \cite{10letter}; finally in \cite{cremonesi-t} they were put in a very compact form:
\begin{subequations}\label{eq:ads7}
	\begin{align}
		&\frac1{\pi \sqrt2} ds^2= 8\sqrt{-\frac \alpha{\ddot \alpha}}ds^2_{{\rm AdS}_7}+ \sqrt{-\frac {\ddot \alpha}\alpha} \left(dz^2 + \frac{\alpha^2}{\dot \alpha^2 - 2 \alpha \ddot \alpha} ds^2_{S^2}\right)\, ;\\
		&B=\pi \left( -z+\frac{\alpha \dot \alpha}{\dot \alpha^2-2 \alpha \ddot \alpha}\right) {\rm vol}_{S^2}\,,\qquad F_2 =  \left(\frac{\ddot \alpha}{162 \pi^2}+ \frac{\pi F_0\alpha \dot \alpha}{\dot \alpha^2-2 \alpha \ddot \alpha}\right) {\rm vol}_{S^2}\,;\\ 
		&e^\phi=2^{5/4}\pi^{5/2} 3^4 \frac{(-\alpha/\ddot \alpha)^{3/4}}{(\dot \alpha^2-2 \alpha \ddot \alpha)^{1/2}}\, .
	\end{align}	
\end{subequations}
$\ddot \alpha = \ddot \alpha(z)$ is a piecewise-linear function on a closed interval $I$ with coordinate $z$. The internal space is topologically an $S^3$; the metric has an $\mathrm{SU}(2)$ isometry acting on the round $S^2$, which realizes the  R-symmetry. There are D8/D6 bound states at the loci $z=z_a$ where $\ddot \alpha$ changes slope (which are copies of $S^2$). Additionally, there may be D6-branes at the endpoints of $I$. 

The correspondence between these AdS$_7$ solutions and the SCFTs is also easy to write down: the $\mu^t_a$, which we defined earlier as the number of boxes in each column of the Young diagram associated to $\mu$, give the slope of the piecewise-linear function $\ddot \alpha$ (see \cite{cremonesi-t} for more details, and especially Fig.~2 there). In this paper we will only need to know that there are stacks of D8-branes realizing each of the factors inside the $\mathrm{S}(\ldots)$ in (\ref{eq:flavor-flav}), with the exception of $\mathrm{U}(f_1^\mathrm{L,R})$, which are realized on a D6-stack. So the theory ${\mathcal T}^N_{\mathrm{SU}(k),0,0}$ is dual to a solution with two stacks with $k$ D6-branes each, while for any other example some of the D6-branes turn into D8-branes. For example for ${\mathcal T}^N_{\mathrm{SU}(k),[k],0}$, we have a single D8 corresponding to the $[k]$ on the left and a stack of $k$ D6-branes corresponding to the $0=[1^k]$ on the right.

In \cite{cremonesi-t}, the $a$ anomaly was also computed from the AdS$_7$ solutions and from the tensor-branch effective theory, finding agreement for any $N$, $\mu_\mathrm{L}$ and $\mu_\mathrm{R}$. This provides a strong check that the solutions (\ref{eq:ads7}) indeed correspond to the SCFTs (\ref{eq:T}).

The $G=\mathrm{SO}(2k)$ can be obtained by suitably orientifolding (\ref{eq:ads7}); this adds two O6-planes to the endpoints of $I$; the holographic anomaly match also works in this case \cite{apruzzi-fazzi}. (An additional possibility is to have an O8 \cite{bah-passias-t,apruzzi-fazzi}; this however corresponds to theories outside the class (\ref{eq:T}).)


\section{The seven-dimensional supergravity theory} 
\label{sec:7dsug}

As mentioned in the introduction, there is already a consistent truncation \cite{prt} connecting each of the AdS$_7$ solutions (\ref{eq:ads7}) to a theory called \emph{minimal gauged supergravity} in seven dimensions, which we will describe shortly. This theory has a single supersymmetric vacuum; this means that it captures only some ``universal'' features common to all the SCFTs (\ref{eq:T}), and cannot be used to describe domain walls connecting them.

Thus for our purposes we should find a reduction that keeps more modes of the internal manifold, and more information about the physics of the SCFTs. An idea already considered in \cite{prt} is that each of the D6- and D8-brane stacks should contribute in seven dimensions a non-abelian vector multiplet, coming from the gauge fields living on them in ten dimensions. This more ambitious consistent truncation was not found in \cite{prt} and will not be found here. 

However, we can try to guess what the seven-dimensional theory would look like. Once we decide the gauge group, we can just couple the appropriate seven-dimensional vector multiplets to the minimal gauged supergravity found in \cite{prt}. 

In the AdS/CFT correspondence, a flavor symmetry in the CFT becomes a gauge symmetry in the bulk. Thus one might at first be tempted to say that the gauge group might be (\ref{eq:flavor-flav}). However, we would like to find a theory that describes several AdS$_7$ solutions at once. Recall then that (\ref{eq:flavor-flav}) is in fact always a subgroup of
\begin{equation}
	\mathrm{SU}(k)\times \mathrm{SU}(k)\,.
\end{equation}
This opens the possibility that we should take this as a gauge group, and that it will be broken to (\ref{eq:flavor-flav}) on its various vacua.

We conclude then that our seven-dimensional theory is minimal gauged supergravity coupled to two ${\rm SU}(k)$ vector multiplets. This theory was worked out in \cite{bergshoeff-koh-sezgin} and recently reviewed for example in \cite{louis-lust,Karndumri:2014hma}.

The fields of minimal gauged supergravity \cite{townsend-vannieuwenhuizen} are
\begin{equation}\label{eq:gravity-mult}
(e^m_\mu\,,\,\psi_\mu^A\,,\,A_\mu^i\,,\,\chi^A\,,\, B_{\mu\nu}\,,\sigma)\,.
\end{equation}
The index $i=1,2,3$ labels three vectors, which realize an $\mathrm{SU}(2)_{{\cal R}_0}$ gauge group; for us this will realize the R-symmetry of the SCFTs. The index $A$ labels the two gravitini and dilatini, transforming in the $\bf{2}$ of $\mathrm{SU}(2)_{{\cal R}_0}$.

Each vector multiplet has the field content
\begin{equation}
(A_{\mu\,R}\,,\,\lambda^A_R\,,\,\phi_{iR})\,;
\end{equation}
the index $R$ runs from 1 to $n=2(k^2-1)$, the number of vector multiplets. When we couple them to the gravity multiplet (\ref{eq:gravity-mult}), the $3n$ scalars $\phi^{iR}$ together parameterize a moduli space
\begin{equation}\label{eq:grassmannian}
	\frac{{\rm SO}(3,n)}{{\rm SO}(3)\times {\rm SO}(n)}\,.
\end{equation}
We can parameterize this space with a coset representative $L^I{}_J\in \mathrm{SO}(3,n)$, where the index $I=(i,R)$ goes from 1 to $3+n$.

In general we could then gauge any $3+n$-dimensional subgroup of $\mathrm{SO}(3+n)$ whose structure constants $f^L{}_{IJ}$ satisfy the ``linear constraint'' which imposes that $f_{IJK}\equiv f^L{}_{IJ}\eta_{KL}$ are totally antisymmetric.  We will not write the Lagrangian here; it can be found in \cite[(2.11)]{louis-lust}. All we will need is the scalar potential
\begin{align}\label{eq:V}
V&=\frac14 e^{-\sigma} \left(
C^{iR}C_{iR}-\frac19 C^2
\right) + 16 h^2 e^{4 \sigma}-\frac{4\sqrt2}{3}\, h\, e^{	3 \sigma/2}C \,,
\end{align}
and the fermionic supersymmetry transformations in absence of gauge fields
\begin{align}\label{eq:susy}
	\delta\psi_\mu&= 2 D_\mu \epsilon - \frac{\sqrt{2}}{30}e^{-\sigma/2} C \gamma_\mu \epsilon - \frac45 h e^{2 \sigma} \gamma_\mu \epsilon \,, \nonumber\\	
	\delta \chi&= - \frac12 \gamma^{\mu} \partial_\mu \sigma \epsilon+ \frac{\sqrt{2}}{30} e^{-\sigma/2}C \epsilon - \frac{16}{5} e^{2 \sigma} h \epsilon \,,\\
	\delta \lambda^R &= i \gamma^\mu P^{iR}_{\mu} \sigma^i \epsilon- \frac{i}{\sqrt2} e^{-\sigma/2} C^{iR} \sigma^i \epsilon \,. \nonumber
\end{align}
(We have suppressed the R-symmetry indices $A$ here, and we will do so from now on.) Here $h$ is a topological mass term, which is necessary in order to find supersymmetric AdS$_7$ vacua \cite{Karndumri:2014hma,prt}. We have also defined\footnote{We follow the formalism in \cite{bergshoeff-jong-sezgin}, where the first index of $L^I{}_J$ is split in $(i,R)$ and both $i$ and $R$ are lowered with $\delta$'s, while the index $J$ is raised and lowered with an $\eta$.} 
\begin{equation}\label{eq:CC}
\begin{split}
&C=-\frac{1}{\sqrt{2}}\,f_{IJK}{L^{I}}_i\,{L^{J}}_j\,{L^{K}}_k\,\epsilon^{ijk} \,, \\ &C_{iR}=\frac{1}{\sqrt{2}}\,f_{IJK}{L^{I}}_j\,{L^{J}}_k\,{L^{K}}_R\,\epsilon^{ijk}\,,
\\
&P_\mu^{iR}= L^{IR}\left(
\delta^{K}_{I}\,\partial_\mu\,+\, {f_{IJ}}^K\,A^{J}_{\mu}
\right)L_K^{i}\,.
\end{split}	
\end{equation}

Putting together the bulk duals of R-symmetry and flavor symmetry, we need\footnote{This is consistent with a general finding by \cite{louis-lust}, which says that supersymmetric AdS$_7$ vacua exist in these theories only if the gauge group is of the form $G_0\times H$, with $G_0 \supset \mathrm{SU}(2)$ and $H$ compact and semisimple.}
\begin{equation}\label{eq:gauge-group}
	G={\rm SU}(2)_{{\mathcal R}_0}\times {\rm SU}(k)\times {\rm SU}(k)\,.
\end{equation}
Thus the structure constants will split as
\begin{equation}\label{eq:s-c}
	f_{IJK}=\left\{ g_3\epsilon_{ijk}\,,\ g_\mathrm{L}f_{rst}\,,\ g_\mathrm{R} f_{\hat r \hat s \hat t} \right\}
\end{equation}
where now $f_{rst}$  and $f_{\hat r \hat s \hat t}$  are the structure constants of the two copies of  $\mathrm{SU}(k)$; both $r$ and $\hat r$ indices go from 1 to $k^2-1$. We are not venturing to offer an identification of the coupling constants $g_3$, $g_\mathrm{L,R}$ we just introduced (and of the topological mass $h$ we saw earlier), again because we do not have an uplift procedure that explicitly takes our theory to IIA supergravity. In \cite{prt} it was found that the uplift procedure found there for the case without vector multiplets required $g_3 = 2 \sqrt2 h$, but this might conceivably get modified for the present case with vectors. In section \ref{sub:cc} we will determine at least one relation among the coupling constants, by using holography.


\section{Vacua} 
\label{sec:vacua}

Having guessed the seven-dimensional supergravity, we now need to somehow come up with an Ansatz to find vacua that can plausibly represent the AdS$_7$ solutions (\ref{eq:ads7}) upon reduction. Our guiding principles will be that we should find vacua that are
\begin{itemize}
	\item in one-to-one correspondence with a choice of two Young diagrams, which are the main data in the SCFTs (\ref{eq:T}) and their AdS$_7$ duals;
	\item on which the residual gauge symmetries reproduce (\ref{eq:flavor-flav}).
\end{itemize}

In fact, to simplify the problem, for the time being we will look for vacua that are determined by the choice of a \emph{single} Young diagram $\mu_\mathrm{L}\equiv \mu$, and where the second copy of $\mathrm{SU}(k)$ in the gauge group is unbroken. So in this section all the fields in the second copy of the $\mathrm{SU}(k)$ vector multiplet will be set to zero.
We will come back to the general case in section \ref{sec:double}. 

\subsection{The Ansatz} 
\label{sub:ans}

It is natural to think that the $\mathrm{su}(2)$ representation (\ref{eq:red-su2}) should play a role: it is naturally associated with the data of the theory, and its stabilizer gives automatically $\mathrm{S}\left(\Pi_a \mathrm{U}(f_a)\right)$, thus reproducing the left half of (\ref{eq:flavor-flav}). Thus we will simply assume
\begin{equation}\label{eq:phii}
	\phi^i = \psi \sigma^i\,.
\end{equation}
(One might try to put a different number $\psi_a$ in front of each block $\sigma^i_a$, but the vacuum equations quickly impose that all the $\psi_a$ are equal.) It might look like the R-symmetry $\mathrm{SU}(2)_{{\mathcal R}_0}$ is broken, because the three $\phi^i$ are different; but this difference can be reabsorbed in the $\mathrm{SU}(2)\subset \mathrm{SU}(k)$ action inside the gauge group defined by the $\sigma^i$. In other words, in this Ansatz the R-symmetry is realized as the diagonal $\mathrm{SU}(2)_{{\mathcal R}}$  of the original $\mathrm{SU}(2)_{\mathcal{R}_0}$ and of an $\mathrm{SU}(2)$ subgroup of the rest of the gauge group.

Moreover, once we expand the $\phi^i$ on a basis of generators $T^r_\mathrm{f}$ of the gauge algebra (in the fundamental representation), we have 
\begin{equation}
	\phi^i= \phi^i_r T^r_\mathrm{f}\,;
\end{equation}
 the matrix $\phi^i_r$ has the right structure to be one of the blocks of the scalars $L_{IJ}.$ 
We will normalize $T^r_\mathrm{f}$ such that $\mathrm{Tr}(T^r_\mathrm{f} T^s_\mathrm{f}) =-\delta^{rs}$.
Recall indeed that the indices $I,J$ decompose naturally as $(i,r)$; so $\phi^i_r$ might be related to the blocks $L_{ir}$ or $L_{sj}$. It appears natural to use the quotient in (\ref{eq:grassmannian}) to set the blocks $L_{ij}$ and $L_{rs}$ to zero. This leads to a matrix 
\begin{equation}
	\left( \begin{array}{cc}
		 0 & \phi^i_r \\
		\phi^s_j & 0 
	\end{array}\right)\,
\end{equation}
where $\phi^s_j = \phi^j_s$. This is in fact an element of the Lie algebra $\mathrm{so}(3,n)$, so it looks promising. The $L^I{}_J$ are in the group $\mathrm{SO}(3,n)$, but this is easily fixed by inserting an exponential. So we end up with
\begin{equation}\label{eq:Lans}
	L^I{}_J= (L^i{}_J, L^r{}_J) =\exp \left[ \begin{array}{cc}
		 0 & \phi^i_r \\
		\phi^s_j & 0 
	\end{array}\right]\,.
\end{equation}

Encouragingly, a very particular case of the Ansatz (\ref{eq:Lans}) was considered in \cite{Karndumri:2014hma}, where it indeed led to a new vacuum. That paper considered minimal gauged supergravity coupled to three vector multiplets, with an $\mathrm{SU}(2)_{{\mathcal R}_0}\times \mathrm{SU}(2)$ gauge group rather than (\ref{eq:gauge-group}); so it is more or less a particular case of the theory in our paper, if we take $k=2$ and leave the second $\mathrm{SU}(k=2)$ factor in (\ref{eq:gauge-group}). For $k=2$, the only non-trivial partition that we can consider is {\tiny $\yng(2)$}. In this case $\phi^i_r$ is simply proportional to $\delta^i_r$; (\ref{eq:Lans}) then becomes the Ansatz in \cite[(3.1)]{Karndumri:2014hma}, which was found there to lead to a supersymmetric vacuum.


\subsection{Finding vacua} 
\label{sub:find-vacua}

All this sounds encouraging; let us now see if we can indeed find vacua with the Ansatz (\ref{eq:Lans}). We first need to compute the exponential in (\ref{eq:Lans}). Already at quadratic order we need to compute $\phi^i_r \phi^j_r$ and $\phi^s_k \phi^t_k$, which we will now proceed to do.

From (\ref{eq:phii}), (\ref{eq:red-su2}) we have
\begin{equation}\label{eq:pp}
\begin{split}
	\phi^i_r \phi^j_r &= -{\rm Tr}(\phi^i_r T^r_\mathrm{f} \phi^j_s T^s_\mathrm{f})= 
	-\mathrm{Tr}\left(\phi^i \phi^j \right)=\psi^2 \sum_a \kappa_{a}^2 \delta^{ij}= \alpha^2 \delta^{ij} \, ,\\
	& \alpha^2 \equiv \psi^2 \kappa^2 \, ,\qquad \kappa^2 \equiv \sum_a \kappa_a^2\,.
\end{split}
\end{equation}
On the other hand, $P^{st}\equiv \phi^s_j \phi^t_j$ is a little more subtle. This can have rank at most 3, and so in particular it cannot be proportional to the identity $\delta^{st}$. However, using (\ref{eq:pp}) we see that it is proportional to a projector:
\begin{equation}\label{eq:PP}
	P^{rs} P^{st} = \phi^r_j \phi^s_j \phi^s_k \phi^t_k = \alpha^2 \phi^r_j \delta^{jk}\phi^t_k= \alpha^2 P^{rt}\,. 
\end{equation}

With (\ref{eq:pp}) and (\ref{eq:PP}), the exponential in (\ref{eq:Lans}) can be resummed and gives
\begin{equation}
	L^I{}_J=\left(\begin{array}{cc}
		\cosh \alpha \delta^{ij} &  \frac{\sinh \alpha}\alpha \phi^i_r\\
		\frac{\sinh \alpha}\alpha  \phi^s_j & \delta^{rs} + \frac{\cosh \alpha -1}{\alpha^2} P^{rs}  
	\end{array}\right)\,.
\end{equation}

We now have to check whether this leads to a supersymmetric vacuum. The quickest way is to use the BPS equations, which consist in setting to zero the fermionic transformations laws (\ref{eq:susy}). On a vacuum, all scalars are constant; then $\delta \chi = 0 = \delta \lambda^r$ give 
\begin{equation}\label{eq:Cir0}
	C_{ir}=0 \, ,\qquad C= 48 \sqrt2 h e^{5/2 \sigma}\,.
\end{equation}

We can compute $C$ and $C_{ir}$ from (\ref{eq:phii}), (\ref{eq:red-su2}), (\ref{eq:su2-rep}):
\begin{subequations}
	\begin{align}\label{eq:C-ans}
	C&=-3\sqrt{2}\left( -g_3 \cosh(\alpha)^3+\frac{g_{\mathrm{L}}}{\kappa}\sinh(\alpha)^3 \right)\,,\\
	C^{ir}&=\frac{\sqrt2}\alpha \cosh \alpha \sinh \alpha \left(-g_3\cosh \alpha + \frac{g_{\mathrm{L}}}{\kappa}\sinh\alpha\right)\phi^{i}_r\,.
	\end{align}	
\end{subequations}
Imposing (\ref{eq:Cir0}) then results in 
\begin{equation}
\label{eq:vacuumScalars}
\tanh(\psi\kappa)=\dfrac{\kappa\,g_3}{g_{\mathrm{L}}}\,,\qquad e^{\tfrac{5\sigma}{2}}=\dfrac{g_3\,g_{\mathrm{L}}}{16\,h\,\sqrt{g_{\mathrm{L}}^2-g_3^2\,\kappa^2}}\,.
\end{equation}

Thus we have succeeded in finding a BPS vacuum for each choice of partition $\mu_\mathrm{L}$. Let us summarize it: the vector multiplet scalars are given by (\ref{eq:phii}), with $\kappa^2=\sum_a \kappa^2_a$ and (\ref{eq:red-su2}) the reducible $\mathrm{su}(2)$ representation associated to $\mu_\mathrm{L}$.  $\psi$ and the gravity multiplet scalar $\sigma$ are determined in (\ref{eq:vacuumScalars}).

The vacua we found are in one-to-one correspondence with a partition $\mu_\mathrm{L}$, as expected. (Recall we have kept $\mu_\mathrm{R}=0$ in this section; we will allow $\mu_\mathrm{R}$ to be nontrivial as well in section \ref{sec:double}.) More precisely, the non-abelian nature of the Ansatz  (\ref{eq:phii}) for the scalars,  and in particular the appearance of a reducible $\mathrm{SU}(2)$ representation (\ref{eq:red-su2}), suggest a Myers-like effect \cite{myers} in which the D6-branes of the trivial vacuum $\phi^i=0$ expand into spherical D8-branes in the internal directions. It was already widely suspected that such an interpretation would be possible for the IIA AdS$_7$ solutions of \cite{afrt,10letter,cremonesi-t}.

These are encouraging signs that these vacua represent the AdS$_7$ solutions (\ref{eq:ads7}) of type IIA. In the next two subsections we will perform some simple checks of this picture. In section \ref{sub:cc} we will consider the cosmological constant in these vacua, comparing it with the one in ten dimensions. In section \ref{sub:mass} we will compute the masses of the scalars around vacua, and consequently the dual operator dimensions.

It would also be possible to look for non-supersymmetric vacua with the same Ansatz. We know that these exist, since one exists already in the minimal theory with no vectors \cite{campos-ferretti-larsson-martelli-nilsson}; given the universal lift of \cite{prt}, in fact we even know that every supersymmetric solution has a non-supersymmetric twin. These were given a CFT interpretation in \cite{apruzzi-dibitetto-tizzano}, but were later found to be unstable within the larger theory with abelian vectors \cite{danielsson-dibitetto-vargas-swamp} (which should represent transverse D6-brane motions rather than the Myers effect described in this paper). Given this instability, we have not analyzed such vacua.


\subsection{Cosmological constant} 
\label{sub:cc}

The cosmological constant on our vacua can be computed from (\ref{eq:V}), which gives $V= -240 e^{4 \sigma}h^2$. In terms of the cosmological constant $V_0$ for the trivial vacuum we get
\begin{equation}\label{eq:cc-7d}
	\left(\frac{V_{\mu_\mathrm{L}}}{V_0}\right)^{5/4} = \frac1{1-\kappa^2\frac{g_3^2}{g_\mathrm{L}^2}}\,.
\end{equation}
We would like to compare this with the computation performed in \cite{cremonesi-t} directly in IIA supergravity. That result was successfully matched there with a field theory computation in the large $N$ limit. More precisely, if one makes $N$ large and nothing else, the D8-branes in the gravity solution become smaller and smaller, ending up with a solution with only D6-branes (the dual of the ${\mathcal T}^N_{\mathrm{SU}(k),0,0}$ theory). To get a more interesting match, one can also make large the D6-charges of the D8s, which are proportional to their radii. In the language of this paper, this means that the dimensions $d_a$ of the blocks in (\ref{eq:red-su2}) are large. So the limit where the holographic match in \cite{cremonesi-t} is most interesting is
\begin{equation}\label{eq:hol-limit}
	N \to  \infty \, ,\qquad d_a \to \infty \, ,\qquad d_a/N\equiv \delta_a \text{ finite}.
\end{equation}
 So we can approximate (\ref{eq:su2-rep}) as $\kappa_a^2\sim \frac{d_a^3}{12}$, and $\kappa^2 \sim \sum_a \frac{d_a^3}{12}$. Recalling (\ref{eq:fa-blocks}), we can rewrite this as 
\begin{equation}\label{eq:kappa-approx}
	\kappa^2 \sim \frac1{12} \sum_a a^3 f_a\,.
\end{equation}
The regime (\ref{eq:hol-limit}) is now the one where the $f_a$ are non-zero only for large $a$.

On the other hand, after some massaging  the expression for the $a$ anomaly given in \cite[(3.15)--(3.16)]{cremonesi-t} can be rewritten as
\begin{equation}\label{eq:crem}
	a_{\mu_\mathrm{L}} = N^3\frac{k^2}{12} - N \frac k6  \sum_a a^3 f_a + \ldots\,,
\end{equation}
where the $\ldots$ denote terms of order $N^0$ and $N^{-1}$. $a$ is in fact proportional to $L_\mathrm{AdS}^5$, which is in turn proportional to $V^{-5/2}$. So we can rephrase (\ref{eq:crem}) as
\begin{equation}\label{eq:cc-10d-approx}
	\left(\frac{V_{\mu_\mathrm{L}}}{V_0}\right)^{-5/2} = 1 - 2 N^{-2} k^{-1} \sum_a a^3 f_a + \ldots\,.
\end{equation}
Taking into account (\ref{eq:kappa-approx}), this matches the behavior we observed in (\ref{eq:cc-7d}), if we identify
\begin{equation}\label{eq:g3lift}
	\frac{g_3^2}{g_\mathrm{L}^2} = \frac1{N k^2}\,.
\end{equation}
As commented at the end of section \ref{sec:7dsug}, so far we had not ventured to identify the parameters of the seven-dimensional theory with those of IIA, because we have no consistent truncation procedure. 

So we managed to match the structure of (\ref{eq:crem}) or (\ref{eq:cc-10d-approx}) with our seven-dimensional supergravity results. Let us look a little more closely. The two terms of (\ref{eq:crem}) are both of order $N^5$ in the limit (\ref{eq:hol-limit}): from (\ref{eq:afak}) we see $k\sim N \delta$, and from (\ref{eq:kappa-approx}) we see $\kappa^2 \sim N^3 \delta$, where $\delta$ is a typical $\delta_a$ as defined in (\ref{eq:hol-limit}). The terms $\ldots$ in (\ref{eq:crem}) in fact also scale like $N^5$ in the limit (\ref{eq:hol-limit}), even though they are superficially of order $N^0$ and $N^{-1}$. These terms were in fact considered in \cite{cremonesi-t}, and they also matched the field theory computation perfectly. 

So one might want to recover these further $\ldots$ terms in (\ref{eq:crem}) or (\ref{eq:cc-10d-approx}) as well. This does not work; but with a little thought one sees why. The vectors multiplets we have added to our pure minimal supergravity in seven dimensions have the usual quadratic action. But in ten dimensions they originate from the brane action, which is not quadratic. For this reason, as we anticipated in the introduction, a IIA reduction can never literally reproduce our seven-dimensional gauged supergravity. For a perfect match, one should improve our vector multiplet action by adding higher-derivative terms, something which is currently beyond the state of the art. 

For this reason, a perfect quantitative match between the cosmological constant as computed in IIA and in our seven-dimensional supergravity can only be obtained when the vev's of the vector multiplet fields are not too large. It is natural to interpret this as saying that the $\delta_a$ in (\ref{eq:hol-limit}) should not be too large (even if the $d_a = \delta_a N$ are large). Under this condition, the $\ldots$ terms in (\ref{eq:crem}), (\ref{eq:cc-10d-approx}) are in fact subdominant. Thus we obtain a match in the regime where our approach can be quantitatively justified. 

In retrospect, it is quite impressive that we still obtain a qualitative match with ten dimensions even beyond this regime, in the sense that we obtain all the vacua expected from ten dimensions. In the next section we will see that even the RG flows between these vacua are in qualitative agreement with expectations, this time from field theory.


\subsection{Masses and dimensions} 
\label{sub:mass}

We now perform another routine computation: the scalar masses around our vacua.

As a warm-up, we notice that it is particularly easy to compute the masses for the dilaton $\sigma$, and for the particular direction in the $L^I{}_J$ space associated to $\psi$, which corresponds to taking $\delta \phi^i \propto \phi^i$. In that case, we can simply rely on the formulas of section \ref{sub:find-vacua}, obtaining
\begin{equation}
	m^2_\sigma=\frac45 \partial_\sigma^2 V = -8 \left(-\frac V{15}\right)=  - \frac 8{L^2_\mathrm{AdS}}
	\, ,\qquad
	m^2_\psi=\frac{1}{3\kappa^2} \partial_\psi^2 V = 40 \left(-\frac V{15}\right) = \frac{40}{L^2_\mathrm{AdS}}\,.
\end{equation}
(The factors $4/5$ and $1/3\kappa^2$ are included to normalize the scalars canonically in the Lagrangian). As expected for a BPS solution, the two masses satisfies the unitarity bound $m^2\,>\,-\frac9{L_\mathrm{AdS}}$. From the usual holographic relation $m^2 L^2_\mathrm{AdS}= \Delta (\Delta-6)$ one reads off the conformal dimensions of the dual SCFT operators $\Delta=4$ and $10$. 

There are many other scalars, and we may in particular wonder about the remaining $3(k^2-1)-1$ scalars in the active $\mathrm{SU}(k)_\mathrm{L}$ vector multiplet. This means we have to consider a more general fluctuation $\delta \phi^i$, not proportional to $\phi^i$. To analyze the masses of such fluctuations, we have to extend a bit our computations from section \ref{sub:find-vacua}. 

First of all we have to be more precise about how the $\delta \phi^i$ appear in the scalar fluctuations. The scalars $L^I{}_J$ live in the coset (\ref{eq:grassmannian}); in particular they are elements of $\mathrm{SO}(3,3+n)$. Then $\delta L L^{-1}\equiv \delta \phi$ is in the Lie algebra $\mathrm{so}(3,3+n)$. We can now parameterize fluctuations with a $\delta \phi	$ that has no generators in the subalgebra $\mathrm{so}(3)\oplus \mathrm{so}(3+n)$. In other words we have
\begin{equation}
	\delta L^I{}_J = \left(\begin{array}{cc}
		0 & \delta \phi^i{}_s \\ \delta \phi^r{}_j &  0 
	\end{array}\right) 
	\left(\begin{array}{cc}
		L^j{}_k & L^j{}_t \\ L^s{}_k & L^s{}_t 
	\end{array}\right)\,.
\end{equation} 
(This approach was also followed in \cite[Sec.~2.3]{louis-lust}.) We can now compute the variations 
\begin{equation}\label{eq:deltaC}
	\delta C^{ir} = C^i_j \delta \phi^{lr} + 2 C^{ijrs} \delta \phi_{js} \, ,\qquad \delta C = -3 C^{ir} \delta \phi_{ir}\,,
\end{equation} 
where
\begin{equation}
	C^{i}_l = \epsilon^{ijk} f_{IJK} L^I{}_j L^J{}_k L^K{}_l \, ,\qquad C^{ijrs} = \epsilon^{ijk}f_{IJK} L^{Ir} L^{Js} L^K{}_k\,.
\end{equation}
$C^{ir}$ vanishes on the vacuum; from (\ref{eq:deltaC}) we also read that $\delta C=0$. The second variation of the potential then reads
\begin{equation}
	\delta^2 V = \frac12 e^{-\sigma} \left( \delta C^{ir} \delta C_{ir} -\frac19 C \delta^2 C\right) - \frac{4\sqrt2}3 h e^{3 \sigma/2} \delta^2 C\,.
\end{equation}
A lengthy computation results in  
\begin{equation}\label{eq:delta2V}
	\delta^2 V = -8\mathrm{Tr}\left( \delta \phi^i \delta \phi^i -2 [\sigma^i, \delta \phi^j][\sigma^i, \delta \phi^j] + 2 [\sigma^i, \delta \phi^j][\sigma^j, \delta \phi^i] + [\sigma^i, \sigma^j][ \delta \phi^i, \delta \phi^j]\right)\,,
\end{equation} 
where $\sigma^i$ denotes the reducible representation (\ref{eq:red-su2}) of su(2), to which the scalars $\phi^i$ were taken proportional in (\ref{eq:phii}). To normalize fields canonically, one also has to evaluate the kinetic term in the Lagrangian, which reads more simply $\mathrm{Tr}(\partial_\mu \delta \phi^i \partial^\mu \delta \phi^i)$. 

Define now
\begin{equation}\label{eq:ji-def}
	[\sigma^i, T^\mathrm{f}_r]\equiv j^i_{rs} T^\mathrm{f}_s\,,
\end{equation}
in terms of the basis $T^\mathrm{f}_r$ of the Lie algebra $\mathrm{su}(k)$. In terms of this definition, we can write the mass matrix (with canonically normalized scalars) as
\begin{equation}\label{eq:mass-matrix}
	M^{ij}_{\alpha \beta}= -8\left( \delta^{ij}(1 + 2 j^k j^k) - 2 j^{(i} j^{j)} \right)_{\alpha \beta}\,.
\end{equation} 
The $\sigma^i$ satisfy the $\mathrm{su}(2)$ algebra. By the Jacobi identity, the $j^i_{rs}$ also satisfy the same algebra: $j^i_{rt}j^j_{ts}-j^j_{rt}j^i_{ts}= \epsilon^{ijk} j^k_{rs}.$ In other words, the $j^i$ form an su(2) representation of dimension $k^2-1$. This representation depends on $k$ and on our choice of block dimensions in (\ref{eq:red-su2}), which are the $d_a$ of the Young diagrams. For example, if $k=2$ and we take the $\sigma^i$ to have a single block (corresponding to $\mu=[2]$), the $j^i$ simply form the $l=1$ representation of su(2). More generally, the representation of the $j^i$ is reducible: it contains several values of $l$. For example, if we take the $\sigma^i$ to be a single dimension $k$ block, corresponding to $\mu=[k]$, the $j^i$ are the direct sum of representations of dimensions $1$, $3$, \ldots $2k+1$. The fully general rule is this: the representation $j^i$ is the reducible representation 
\begin{align}
	\left( \mathbf{d_1}\oplus \mathbf{d_2} \oplus \ldots \right) &\otimes 
	\left( \mathbf{d_1}\oplus \mathbf{d_2} \oplus \ldots \right) \nonumber\\
	=&\oplus_a (\mathbf{2 d_a -1}\oplus \mathbf{2d_a-3}\oplus \ldots \mathbf{1})\oplus \label{eq:ji}\\
	&2\oplus_{a>b} (\mathbf{d_a + d_b-1 } \oplus \mathbf{d_a + d_b-3 }\oplus \ldots \oplus \mathbf{d_a-d_b+1})\,, \nonumber
\end{align}
subtracting one singlet $\mathbf{1}$ from the result. Here we denote irreducible su(2) representations by their dimension. 

Now we can evaluate the mass matrix (\ref{eq:mass-matrix}). The term $j^kj^k$ is a Casimir invariant; the matrix $N^{ij}_{\alpha \beta} \equiv (j^{(i} j^{j)})_{\alpha \beta}$ is more complicated, but it can be seen to have (on a representation of spin $l$) eigenvalues $\{-l(l+1)+\frac12,\frac12 (l+1),-\frac l2\}$, with multiplicities $2l+1$ (or 0 if $l=0$), $2l-1$, $2l+3$ respectively. With this information one can obtain the masses $m^2$ as eigenvalues of (\ref{eq:mass-matrix}), and the corresponding operator dimensions again with the formula $m^2 L^2_\mathrm{AdS}= \Delta (\Delta-6)$. The results of this analysis are detailed in Table \ref{tab:dim}. We list them in terms of $\mathrm{SU}(2)_{\mathcal R}$ representations, which as we mentioned below (\ref{eq:phii}) is the diagonal of the original $\mathrm{SU}(2)_{{\mathcal R}_0}$ and of the $\mathrm{SU}(2) \subset \mathrm{SU}(k)$ defined by the $\sigma^i$. The results of Table \ref{tab:dim} should be applied to all the irreducible $\mathrm{SU}(2)$ representations contained in (\ref{eq:ji}).

\begin{table}[t]
	\centering
\begin{tabular}{|c|c|}
\hline
$ \Delta $ & $\mathrm{SU}(2)_{\mathcal R}$ rep. \\
\hline\hline
6 & \textbf{d} \\
4l+6 =2d+4 & \textbf{d-2} \\
4l+4 =2d+2 & \textbf{d+2} \\
\hline
\end{tabular}
\caption{\small Operator dimensions, and their R-symmetry representation. The $\Delta=6$ in the first line is absent in the singlet case, $\mathbf{d=1}$. The second line is only present for $d>2$.}
\label{tab:dim}
\end{table}

As an example, if we consider $k=2$ with a single block, corresponding to the partition $\mu=[2]$, as we mentioned earlier (\ref{eq:ji}) gives us a single triplet, $\mathbf{d=3}$; Table \ref{tab:dim} then produces an operator with $\Delta=6$ in the $\mathbf{3}$, one with $\Delta=10$ in the $\mathbf{1}$, and one with $\Delta=8$ in the $\mathbf{5}$. This agrees with \cite[Sec.~3.1]{Karndumri:2014hma}. 

The presence of marginal operators ($\Delta=6$) deserves some comment. They would seem to suggest the presence of deformations for our vacua. This seems to disagree with general arguments \cite{louis-lust,cordova-dumitrescu-intriligator-def} forbidding supersymmetric deformations for AdS$_7$ vacua (or their CFT$_6$ duals), and in fact with the classification of type II AdS$_7$ vacua \cite{afrt,10letter,cremonesi-t} that we want to reproduce. However, recall that part of our gauge group has been broken, in the pattern $\mathrm{SU}(k)_\mathrm{L}\to \mathrm{S}(\Pi_a \mathrm{U}(f_a^\mathrm{L}))$. The broken gauge vectors have obtained a mass, and thus have eaten some scalars. One can show that the number of such gauge vectors is exactly equal to the $\Delta=6$ operators from the first line of Table \ref{tab:dim}. 

There are additional $\Delta=6$ operators, coming from the last line of Table \ref{tab:dim} for $l=1/2$ (or $\mathbf{d}=2$). However, so far we have looked at all deformations, without examining whether they are supersymmetric or not. Following \cite{louis-lust}, supersymmetric deformations turn out to be those that satisfy
\begin{equation}\label{eq:susy-def}
	\delta \phi^i = \epsilon^{ijk} [\sigma^i, \delta \phi^k]\,.
\end{equation}
In terms of the $j^i$ in (\ref{eq:ji-def}), this is equivalent to finding eigenvectors of the matrix $E^{ij}_{\alpha \beta} \equiv \epsilon^{ijk} j^k_{\alpha \beta}$ with eigenvalue 1. This matrix commutes with the earlier matrix $N^{ij}_{\alpha \beta} \equiv (j^{(i} j^{j)})_{\alpha \beta}$, which appeared in the mass matrix (\ref{eq:mass-matrix}); so they are simultaneously diagonalizable. The eigenspace of $E$ with eigenvalue 1 is precisely the first line of Table \ref{tab:dim}; all the others give different eigenvalues, so do not satisfy condition (\ref{eq:susy-def}) and do not correspond to supersymmetric deformations. Thus we have no contradiction with the results in \cite{louis-lust,cordova-dumitrescu-intriligator-def}.



\section{Domain walls} 
\label{sec:dw}


The AdS$_7$ vacua obtained so far can be connected by supersymmetric domain walls corresponding to RG flows in the dual 6D SCFT. Their construction is the subject of this section.

The first order BPS scalar flow will be described as a gradient flow in section \ref{sub:BPSflow}.
In section \ref{sub:bps-solving} we will reduce the BPS equations to a study of Nahm equations. In section \ref{sub:nahm} we will review the literature about those equations, showing that solutions exist exactly when they are expected to exist from the point of view of the field theory duals. 

\subsection{Killing spinors and first order flow}
\label{sub:BPSflow}

By setting to zero the fermionic supersymmetry variations \eqref{eq:susy} we obtain the BPS equations for the scalar fields and the Killing spinor preserved along the flow. We are going to derive their form explicitly in this section.

Let's consider the following Ansatz for the domain walls metric
\begin{equation}\label{eq:DW-metric}
	ds_{7}^2=e^{2A(\rho)}ds^2_{\mathrm{Mink}_6} +e^{2 B(\rho)}d\rho^2 \, ,
\end{equation}
with $\rho$ the radial coordinate corresponding to the direction of the flow, and define a superpotential function
\begin{equation}
\label{W-superp}
	W(\sigma,\phi^{ir})=\frac{\sqrt2}{30} e^{- \sigma/2}C+\frac45 h e^{2 \sigma}\,.
\end{equation}
If all fields have only radial dependence, by imposing the projection $e^{B} \gamma^{\rho} \epsilon= \gamma^1 \epsilon=\epsilon$ the BPS equations yield\footnote{Notice that $\delta C=-3 C^{ir}\delta\phi_{ir}$, thus $\partial_{\phi^i_r}W=-\frac1{5\sqrt2}C^{ir}e^{-\sigma/2}$.}
\begin{align}\label{eq:sigma-Pir}
		\sigma'&=-4 e^B\partial_\sigma W\,,
		\qquad
		P^{ir}_\rho= - 5 e^B \partial_{\phi^i_r}W \,.
\end{align}
The BPS equation obtained from the gravitino variation requires more attention. In the covariant derivative of the Killing spinor, $D_\mu \epsilon = \partial_\mu \epsilon +\frac14 \omega_\mu^{mn}\gamma_{mn} \epsilon+\frac i4  \epsilon^{ijk}Q_{\mu jk}\sigma^i\epsilon$, the term containing 
\begin{equation}
	Q_\mu^{ij}=L^{Ij}(\delta^{KI}\partial_\mu+f_{IJ}^{\ \ K}A_\mu^J)L^{\ i}_K
\end{equation}
would require an additional projection on $\epsilon$ involving $\sigma^i$, that however restricts the number of preserved supersymmetries along the flow. We therefore make the Ansatz, consistent with the solutions considered in the rest of this paper, that $Q_{\mu [ij]}=0$. By setting
\begin{equation}\label{eq:A}
 	A'=e^B W\,,
\end{equation}
the BPS equations obtained from $\delta\psi_\mu$ reduce to
\begin{align}
	\partial_\rho \epsilon & =  \frac12 e^{B} W(\sigma,\phi^i_a) \,, 
	\qquad \partial_{\hat\mu} \epsilon =0 \,,
\end{align}
where $x^{\hat\mu}=\{t,x^i\}$ corresponds to Mink$_6$ coordinates. They can be easily integrated to 
\begin{align}
 	\epsilon(r)&=\exp\left( \frac12 \int^\rho e^{B}W(\sigma(\rho'),\phi(\rho')) d\rho' \right) \eta
\end{align}
for a constant spinor satisfying $\gamma_1 \eta=\eta$, that parametrizes the residual $1/2$-supersymmetry along the flow. 

At each endpoint of the flow, \eqref{eq:sigma-Pir} and \eqref{eq:A} describe an AdS$_7$ vacuum, fully supersymmetric, satisfying the attractor equations and Killing spinor equation
\begin{align}\label{eq:AdS7-attr}
 	\langle \partial_\sigma W \rangle&=0\,,
	\qquad
	\langle\partial_{\phi^i_r}W \rangle = 0\,,
	\qquad
	 D_\mu \epsilon -  \frac12 \langle W\rangle \gamma_\mu \epsilon=0 \,.
\end{align}
The latter is simply the Killing spinor equation for AdS with radius $L=\langle W\rangle^{-1}$, which explains the enhancement of supersymmetry at the vacuum. The AdS$_7$ geometries found in section \ref{sec:vacua} are solutions of \eqref{eq:AdS7-attr}.

Notice finally that, in terms of the superpotential $W(\sigma,\phi^i_r)$, the scalar potential \eqref{eq:V} can be written as
\begin{equation}
 	V=5 \left(- 3 W^2 + 2 \partial_\sigma W^2 +\frac52 \partial_{\phi^i_r}W\partial_{\phi^r_i}W\right) \,.
\end{equation}
It can be easily verified that the BPS flow, expressed as the gradient flow (\ref{eq:sigma-Pir}), (\ref{eq:A}),
implies the second order equations of motion of the scalars of seven-dimensional half-maximal supergravity in absence of gauge fields. The warp factor $e^{2B}$ represents a choice of radial parametrization and thus it is not constrained by the first order flow. We will show in the remaining of this section that, in order to solve the equations for the scalars, a convenient radial parametrization will be necessary, leading to a choice for $e^{2B}$.


\subsection{Solving the BPS equations} 
\label{sub:bps-solving}

In section \ref{sec:vacua} we have found a large set of vacua of seven-dimensional supergravity coupled to vector multiplets, in one-to-one correspondence with a choice of partition $\mu_\mathrm{L}$.
In this section we are going to explicitly construct BPS domain walls connecting two such vacua. 

In the metric (\ref{eq:DW-metric}) we will redefine
\begin{equation}
\label{eq:flow-metric}
 	2B = 2 Q + \sigma\,,
\end{equation}
for later convenience. At $\rho\to \pm \infty$, we will impose that $A\to A_\pm \rho$, where $A_\pm$ are two constants, whereas $Q$ and all the scalars (including $\sigma$) will have to become constants; for example 
\begin{equation}\label{eq:interp}
	\phi^i( - \infty) = \phi^i_{\mu_\mathrm{\mathrm{L}-}}
	\, ,\qquad
	\phi^i( + \infty) = \phi^i_{\mu_\mathrm{\mathrm{L}+}} \,,
\end{equation}
where $\phi^i_\mathrm{\mathrm{\mu_{\mathrm{L}\pm}}}$ will be proportional to the $\mathrm{su}(2)$ representations associated to two partitions $\mu_{\mathrm{L}\pm}$ as in (\ref{eq:red-su2}).
(As we anticipated, in this section we are still keeping $\mu_\mathrm{R}=0$.) The limits $\rho\to \pm \infty$ represent respectively the ultraviolet (UV) and infrared (IR) limits of the RG flow.

We will again find a solution by solving the BPS equations \eqref{eq:sigma-Pir}, (\ref{eq:A}). $C$ and $C_{ir}$ now are different from (\ref{eq:C-ans}) because we no longer assume the $\phi^i$ to be proportional to a reducible su(2) representation as in (\ref{eq:phii}). Moreover, the $P^{ir}$ do not vanish, since the scalars now depend on the radial coordinate $\rho$. We obtain
\begin{subequations}\label{eq:PCCdw}
\begin{align}\label{eq:Pdw}
	P^{ir}_\rho&= -\phi^i_r\dfrac{\sinh \alpha}{\alpha}\partial_\rho\cosh \alpha+\left(\delta_{rs}+\dfrac{\cosh \alpha -1}{\alpha^2}P_{rs}\right)\partial_\rho\left( \dfrac{\sinh \alpha}{\alpha}\phi^i_s \right)\,,\\
	\label{eq:Cirdw}
	C^{ir}&=\sqrt 2\left(-\frac{g_3}\alpha \cosh^2\alpha\sinh \alpha\phi^i_r+ g_\mathrm{L}\dfrac{\sinh^2\alpha}{2\alpha^2}\left(\delta_{rs}+\dfrac{\cosh \alpha-1}{\alpha^2}P_{rs}\right)\left[\phi^j,\ \phi^k\right]_s\epsilon^{ijk}\right)\,,\\
	\label{eq:Cdw}
	C&= \dfrac 1{\sqrt 2}\left(6g_3 \cosh^3 \alpha+g_\mathrm{L} \frac{\sinh^3 \alpha}{\alpha^3} \epsilon^{ijk}\mathrm{Tr}\left(\phi_i\left[\phi_j\,,\ \phi_k\right] \right)\right)\,;	
\end{align}	
\end{subequations}
recall $\alpha$ was defined by $\mathrm{Tr}(\phi^i \phi^j)= -\alpha^2 \delta^{ij}$ back in (\ref{eq:pp}).

We start by imposing $\delta \lambda=0$, which is the second in (\ref{eq:sigma-Pir}). The presence in (\ref{eq:PCCdw}) of the projector $P_{rs}$ acting on the derivative and on the commutator of $\phi$'s makes solving the equation apparently problematic. However, the recurring combination
\begin{equation}
\Pi_{rs}\equiv\delta_{rs}+\dfrac{\cosh \alpha-1}{\alpha^2}P_{rs}
\end{equation}
is an invertible operator:
\begin{equation}
\Pi^{-1}_{rs}=
\delta_{rs}-\dfrac{\cosh \alpha-1}{\alpha^2\cosh \alpha}P_{rs}\,.
\end{equation}
Applying $\Pi^{-1}$ to $\delta \lambda=0$, we get
\begin{equation}\label{eq:eQnahm}
e^{-Q}\partial_\rho\Phi^i= \cosh \alpha\left(-g_3\Phi^{i}+\dfrac{1}{2}\left[ \Phi^j\,,\,\Phi^k \right] \epsilon^{ijk} \right)\, ,\qquad 
\Phi^i = g_\mathrm{L}\phi^i \dfrac{\tanh \alpha}{\alpha}\,.
\end{equation}
(The presence of $e^{-Q}$ is due to the vielbein in $\gamma^\mu P^{ir}_\mu$.) We see that it is useful to fix the radial gauge by setting 
\begin{equation}\label{eq:e-Q}
	e^{-Q} = \cosh \alpha,
\end{equation}
which will be our choice from now on. With this, (\ref{eq:eQnahm}) becomes
\begin{equation}\label{eq:nahm}
\partial_\rho\Phi^i= - g_3\Phi^{i}+\dfrac{1}{2}\left[ \Phi^j\,,\,\Phi^k \right] \epsilon^{ijk} \,.
\end{equation}
This is a variant of the Nahm equation, to which it can be mapped by a change of variables. We will study it in section \ref{sub:nahm}; we will show that there exist solutions which at $\rho \to \pm \infty$ approach two different vacua of the type we found in section \ref{sec:vacua}, as in (\ref{eq:interp}).

For now we turn to the other BPS equations, showing that they can be completely solved once a solution $\Phi^i(\rho)$ (and hence $\alpha(\rho)$) of (\ref{eq:nahm}) has been found. $\sigma$ and $A$ are determined by the first in (\ref{eq:sigma-Pir}) and by (\ref{eq:A}). We replace the commutator in $C$ using (\ref{eq:eQnahm}). We obtain the equations
\begin{subequations}\label{eq:subBPS}
\begin{align}
&\partial_\rho\left(\dfrac{e^{-5\sigma/2}}{\cosh\alpha}\right)+g_3\dfrac{e^{-5\sigma/2}}{\cosh\alpha}-\dfrac{16 h}{\cosh^2\alpha} =0\,, \label{eq:sigma'}\\
& \partial_\rho A \cosh \alpha-\dfrac{1}{5}(g_3-\partial_\rho\cosh \alpha)-\dfrac{4}{5} h e^{5\sigma/2} =0\,. \label{eq:A'}
\end{align}	
\end{subequations}
(\ref{eq:sigma'}) can be immediately solved for $\sigma(\rho)$ analytically by performing an integral:
\begin{equation}
	e^{-\tfrac{5}{2}\sigma}=16h\,e^{-g_3}\cosh \alpha\int_{r_0}^r \dfrac{e^{g_3 y}}{\cosh^2\alpha(y)}dy
\end{equation}
A linear combination of (\ref{eq:subBPS}) then gives
\begin{equation}
\partial_r\left( e^{4A+\sigma/2} \cosh\alpha\right)- g_3  e^{4A+\sigma/2} \cosh\alpha =0
\end{equation}
whose solution is 
\begin{equation}\label{eq:A-dw}
A =  \dfrac{1}{4} \left(g_3 \rho -\log\cosh\alpha-\dfrac{1}{2}\sigma \right) + A_0 \,.
\end{equation}

Thus $A(\rho)$ and $\sigma(\rho)$ are determined by the BPS equations, as promised. One can also check that they obey the appropriate boundary conditions we demanded at the beginning of this section, i.e.~$\sigma$ goes to constants $\sigma_\pm$ and $A$ goes as $A_\pm \rho$ at $\pm \infty$. Moreover, the precise values of $\sigma_\pm$ and $A_\pm$ agree with the values determined for the vacua in section \ref{sec:vacua}. In particular, for the cosmological constants we find
\begin{equation}
\left(\dfrac{V_+}{V_-}\right)^{5/4} =  \dfrac{g_\mathrm{L}^2-\kappa^2_- g^2_3}{g_\mathrm{L}^2-\kappa^2_+ g^2_3}\,,
\end{equation}
in agreement with (\ref{eq:cc-7d}). In the next subsection we will see  $\kappa_+ < \kappa_-$; it then follows $V_+ < V_-$, as expected for a domain wall representing an RG flow.

\subsection{Nahm equations} 
\label{sub:nahm}

We will now review why (\ref{eq:nahm}) have solutions with boundary conditions (\ref{eq:interp}), using results in \cite{kronheimer}. (Those results are also reviewed nicely in \cite{bachas-hoppe-pioline}, where (\ref{eq:nahm}) appears in the context of domain walls for the so-called ${\mathcal N}=1^*$ field theory in four dimensions.)

By the change of variables $\Phi_i = \frac 1{g_3s}T_i$, $s=-\frac12 e^{-g_3 \rho}$, (\ref{eq:nahm}) becomes the classic Nahm equation
\begin{equation}
	\partial_s T_i = -\frac12 \epsilon_{ijk}[T_j,T_k]\,.
\end{equation}
This is encouraging, since this equation is very well-studied; however, for us this transformation will be a bit of a curiosity, since in fact our (\ref{eq:nahm}) has already been studied in \cite{kronheimer} essentially already as it is. (If one wants to make contact with the notation there, one can rescale $\rho=\frac2{g_3}t$, $\Phi^i= -\frac{g_3}2 A^i$.) 

Translated in our language, the main result in \cite{kronheimer} is that the moduli space of solutions to (\ref{eq:nahm}) with boundary conditions (\ref{eq:interp}) is the space\footnote{An intersection between a nilpotent orbit and a Slodowy slice appears in several places in the literature, perhaps most notably as the moduli space of the three-dimensional theory ${\mathcal T}^\lambda_\mu$ in \cite{gaiotto-witten-1}. While a similar description applies to Higgs moduli spaces of six-dimensional theories, it would require using orbits of a group much larger than $\mathrm{SU}(k)$. This is equivalent to the formula (\ref{eq:dimH-red}) \cite[Sec.~2.2]{mekareeya-rudelius-t}, which is in terms of orbits of $\mathrm{SU}(k)$ but does not seem to have the structure (\ref{eq:OS}).}
\begin{equation}\label{eq:OS}
	{\mathcal O}(\mu_{\mathrm{L}-}) \cap {\mathcal S}(\mu_{\mathrm{L}+})\,.
\end{equation}
Here ${\mathcal S}(\mu_{\mathrm{L}-})$ is the so-called \emph{Slodowy slice}:
\begin{equation}
	{\mathcal S}(\mu)\equiv \{ \phi^-_\mu + X \ | \ [X,\phi^+_\mu]=0 \}\,, 
\end{equation}
where $\phi^i_\mu$ give the embedding $\mathrm{su}(2)\to \mathrm{su}(k)$ associated to $\mu$, as we discussed around (\ref{eq:su2}), and $\phi^\pm \equiv \frac12(\phi_1 \pm i \phi_2)$. This space has the property of intersecting ${\mathcal O}(\mu)$ in only one point (namely $\phi^-_\mu$). Moreover, it intersects every orbit ${\mathcal O}_{\mu'}$ such that $\mu > \mu'$. For example, $S_0$ is any matrix, and it then meets every orbit. So (\ref{eq:OS}) is non-empty if and only if
\begin{equation}
	\mu_{\mathrm{L}+} < \mu_{\mathrm{L}-}\,.
\end{equation}
The dimension of (\ref{eq:OS}) is
\begin{equation}
	\frac12\left(\sum_a (\mu_{\mathrm{L}+}^t)^2 - \sum_a (\mu_{\mathrm{L}-}^t)^2\right)\,. 
\end{equation}

Thus the orbit $\mu_{\mathrm{L}+}$, corresponding to the theory in the UV, should be dominated by the orbit $\mu_{\mathrm{L}-}$ corresponding to the theory in the IR. This precisely agrees with our field theory prediction (\ref{eq:RG}). 

We will now give a simple example (taken from \cite{bachas-hoppe-pioline}) where the solution can be found explicitly, and analyze it in more detail. It regards the case when the UV partition $\mu_{\mathrm{L}+}$ is $0$. In that case
\begin{equation}\label{eq:trivial}
	\Phi^i = \frac{g_3}{1+e^{g_3 \rho}}\phi^i_{\mu_{\mathrm{L}-}} \,
\end{equation}
is a solution for any $\mu_\mathrm{L}$. The constant matrices $\phi^i_{\mu_{\mathrm{L}-}}$ form the (reducible) SU(2) representation associated to the partition $\mu_{\mathrm{L}-}$, normalized as in (\ref{eq:su2-rep}). Recalling the change of variable performed in (\ref{eq:eQnahm}), the actual matrices parametrizing the scalar manifold read:
\begin{equation}
\alpha\,=\,\mathrm{arctanh}\left[\frac{g_3\,\kappa_{\mathrm{L}-}}{g_L}\frac{1}{1+e^{g_3 \rho}}\right]\quad \Rightarrow \quad \phi^i\,=\,\frac{1}{\kappa_{\mathrm{L}-}}\mathrm{arctanh}\left[\frac{g_3\,\kappa_{\mathrm{L}-}}{g_\mathrm{L}}\frac{1}{1+e^{g_3 \rho}}\right]\phi^i_{\mu_{\mathrm{L}-}}.
\end{equation}
Using (\ref{eq:subBPS}), it is possible to also compute the dilaton $\sigma$ and the warping $A$, although their explicit expressions are quite involved. 

In the UV limit $\rho\,\rightarrow\,+\infty$,\footnote{In the following analysis, we will set $16\,h\,=\,g_{\mathrm{L}}$; with this choice, the vacuum expectation value of the dilaton is zero for $\mu_{\mathrm{L}}\,=\,[1^6]$.} the metric is asymptotic to 
\begin{equation}
ds^2_7\,\sim\, e^{2\frac{g_3\,\rho}{4}}ds^2_{\mathrm{Mink}_6}\,+\, d \rho^2\,,
\end{equation}
which we recognize as the AdS$_7$ metric of radius $L_+= g_3/4$. The dilaton and the scalar fields  behave as
\begin{equation}
\label{eq:asymptotic+}
\phi^i\,\sim\,e^{-\frac{4\rho}{L_+}}\,\phi^i_{\mu_{\mathrm{L}-}}\,,\quad \sigma\,\sim\,e^{-\frac{4\rho}{L_+}}\,.
\end{equation}
Recall that in general the asymptotic behavior of a scalar behaves as $\delta\varphi\,\approx\,\varphi_{\text{nonnorm}} e^{-(6-\Delta) \rho}\,+\,\varphi_{\text{norm}} e^{-\Delta \rho}$; the first term corresponds to deforming the theory by an operator ${\mathcal O}$ of dimension $\Delta$, while the second contribution is associated to giving a vev to ${\mathcal O}$. In our case, we interpret (\ref{eq:asymptotic+}) as saying that the RG flow is triggered by a vev of two operators both with dimension $\Delta=4$.

There is also a simple modification \cite[(2.17)]{bachas-hoppe-pioline} of (\ref{eq:trivial}), also analytical, which connects the partition $[k/2, k/2]$ to the partition $[k/2 + 1, k/2 - 1]$; in other words, it moves a single block from one row of the diagrams to the next.


\section{Two-tableau generalization} 
\label{sec:double}

We will now study how the previous two sections get modified if one also makes $\mu_\mathrm{R}$ nontrivial. 

\subsection{Two-tableau vacua} 
\label{sub:double-vac}

Our Ansatz in this case is
\begin{equation}
	\phi^i=\left( \begin{matrix}\psi_\mathrm{L}\,\sigma_\mathrm{L}^i &\\ & \psi_\mathrm{R}\,\sigma^i_\mathrm{R} \end{matrix} \right)\,,
\end{equation}
where $\psi_\mathrm{L,R}$ are two numbers, and $\sigma^i$ are two reducible representations of su(2). (Recall (\ref{eq:red-su2}) for the $\mu_\mathrm{R}=0$ case.) We have then
\begin{equation}
	[\sigma^i_\mathrm{L,R},\sigma^j_\mathrm{L,R}]= \epsilon^{ijk}\sigma^k_\mathrm{L,R} \, ,\qquad 
	\mathrm{Tr}(\sigma^i_\mathrm{L,R}\sigma^j_\mathrm{L,R})= \kappa^2_{L,R} \delta^{ij}\,.
\end{equation}
Recalling (\ref{eq:s-c}), $\phi^i=\phi^i_r T^r_\mathrm{f}$ can now be further decomposed as $\phi^i_r T^r + \phi^i_{\hat r} T^{\hat r}$. We have
\begin{equation}\label{eq:alpha-LR}
	\phi^i_r \phi^r_j= \mathrm{Tr}_\mathrm{L}(\phi^i \phi^j) = \psi^2_\mathrm{L} \kappa^2_\mathrm{L} \equiv \alpha^2_\mathrm{L} \delta_{ij}\, ,\qquad
	\phi^i_{\hat r} \phi^{\hat r}_j = \mathrm{Tr}_\mathrm{L}(\phi^i \phi^j) =\psi^2_\mathrm{R} \kappa^2_\mathrm{R} \equiv \alpha^2_\mathrm{R} \delta_{ij}\,.
\end{equation}
Moreover, now we have three different projectors:
\begin{equation}
\phi^i_{r}\,\phi^i_{s}\,=\, P_{r s} \, ,\qquad
 \phi^i_{\hat r}\,\phi^i_{\hat s}\,=\, P_{\hat r \hat s} \, ,\qquad
 \phi^i_{r}\,\phi^i_{\hat s}\,=\,T_{r \hat s}\,.
\end{equation}
They act on the fields as
\begin{align}
&P_{r s}\,\phi_i^{s}\,=\,\alpha_L^2\,\phi^i_{r} \, ,\qquad 
 P_{\hat r \hat s}\,\phi_i^{\hat s}\,=\,\alpha_\mathrm{R}^2\,\phi^i_{\hat r} \\
&T_{r \hat s}\,\phi_i^{\hat s}\,=\,\alpha_\mathrm{R}^2\,\phi^i_r
 \, ,\qquad
  T_{\hat r s}\,\phi_i^{s}\,=\,\alpha_\mathrm{L}^2\,\phi^i_{\hat r}\,, 
\end{align}
and satisfy the relations
\begin{equation}
\begin{split}
P_{r s}\,P^{s t}\,=\, \alpha^2_\mathrm{L}\,P_{r}{}^{t} \, ,\qquad P_{\hat r \hat s}\,P^{\hat s \hat t}\,=\, \alpha^2_\mathrm{R}\,P_{\hat r}{}^{ \hat t}\\
T_{r \hat s}\,T^{\hat s t}\,=\, \alpha^2_\mathrm{R}\,P_{r}{}^{t} 
\, ,\qquad T_{\hat r s}\,T^{s \hat t}\,=\, \alpha^2_\mathrm{L}\,P_{\hat r}{}^{\hat t}\\
P_{r s}\,T^{s \hat t}\,=\, \alpha^2_\mathrm{L}\,T_{r}{}^{\hat t} \, ,\qquad P_{\hat r\hat s}\,T^{\hat s t}\,=\, \alpha^2_\mathrm{R}\,T_{\hat r}{}^{ t}
\end{split}
\end{equation}
At this point we can compute our scalar matrix as:
\begin{equation}\label{eq:L-double}
L^I{}_J= \exp\left( \begin{matrix} 0 & \phi^i_s & \phi^i_{\hat s} \\ \phi^r_j & 0 & 0 \\ \phi^{\hat r}_j & 0 & 0 \end{matrix} \right)\,=\, 
\left( 
\begin{array}{ccc}
\delta^{ij}\,\cosh \alpha  & \dfrac{\sinh \alpha}{\alpha}\,\phi^i_s & \dfrac{\sinh \alpha}{\alpha}\,\phi^i_{\hat s} \\[1.5ex]
\dfrac{\sinh \alpha}{\alpha}\phi^r_j & \delta_{r s}+\dfrac{\cosh \alpha-1}{\alpha^2}\,P_{r s} & \dfrac{\cosh \alpha-1}{\alpha^2}\,T_{r\,\hat s} \\[1.5ex]
 \dfrac{\sinh \alpha}{\alpha}\phi^{\hat r}_j  & \dfrac{\cosh \alpha-1}{\alpha^2}\,T_{\hat r\,s}& \delta_{\hat r\hat s}+\dfrac{\cosh \alpha-1}{\alpha^2}\,P_{\hat r\hat s}
\end{array}
\right)\,,
\end{equation}
with
\begin{equation}
	 \alpha^2\equiv \alpha^2_\mathrm{L}+ \alpha^2_\mathrm{R}\,.
\end{equation}

Now the quantities in (\ref{eq:susy}) are 
\begin{equation}\label{eq:C-double}
\begin{split}
	C\,&=\, 3\sqrt{2}\left( g_3\cosh^3 \alpha\,- 
	\frac{\sinh^3 \alpha}{\alpha^3} \hat C
	 \right) \, ,\qquad 
	\hat C \equiv g_\mathrm{L}\,\psi_\mathrm{L}\,\alpha^2_\mathrm{L}\,+\, g_\mathrm{R}\,\psi_\mathrm{R}\,\alpha^2_\mathrm{R}\,,
	 \\
	C_{i r}\,&=\, \sqrt{2}\frac{\sinh \alpha}\alpha\left( -g_3\,\cosh^2 \alpha\,+\, \dfrac{\sinh \alpha}{\alpha}\dfrac{\cosh \alpha-1}{\alpha^2}\hat C\,+\,\psi_\mathrm{L}\,g_\mathrm{L}\,\dfrac{\sinh \alpha}{\alpha}\, \right)\phi^i_r \,,\\
	C_{i \hat r}\,&=\, \sqrt{2}\frac{\sinh \alpha}\alpha\left( -g_3\,\cosh^2 \alpha\,+\, \dfrac{\sinh \alpha}{\alpha}\dfrac{\cosh \alpha-1}{\alpha^2}\hat C\,+\,\psi_\mathrm{R}\,g_\mathrm{R}\,\dfrac{\sinh \alpha}{\alpha}\, \right)\phi^i_{\hat r}\,.
\end{split}	
\end{equation}

Just like in (\ref{eq:Cir0}), we need to impose $C_{i r}=0$, which now reads $C_{i r}=C_{i \hat r}=0$. From (\ref{eq:C-double}) we then see
\begin{equation}
g_\mathrm{L}\,\psi_\mathrm{L}\,=\, g_\mathrm{R}\,\psi_\mathrm{R}\,.
\end{equation}
Thus we can parametrise everything in terms of a single constant:
\begin{equation}
\psi\equiv \psi_\mathrm{L} \, ,\qquad \psi_\mathrm{R}\,=\, \dfrac{g_\mathrm{L}}{g_\mathrm{R}}\,\psi\,.
\end{equation}
We also define
\begin{equation}
\beta^2\,=\, \dfrac{\kappa_\mathrm{L}^2}{g_\mathrm{L}^2}\,+\,\dfrac{\kappa_\mathrm{R}^2}{g_\mathrm{R}^2} \qquad \Rightarrow\qquad \alpha\,=\, \psi\,\beta\,g_\mathrm{L}\,.
\end{equation}

Going back to $C_{i r}=C_{i \hat r}=0$, we now get
\begin{equation}
 g_3 \cosh(\psi\, \beta\,g_\mathrm{L})=\dfrac{\sinh(\psi\,\beta\,g_\mathrm{L})}{\beta}
\end{equation}
and thus finally we get
\begin{equation}
\psi_\mathrm{L}\,=\, \dfrac{1}{g_\mathrm{L}\,\beta}\mathrm{arctanh}(g_3\,\beta)
\, ,\qquad
\psi_\mathrm{R}\,=\, \dfrac{1}{g_\mathrm{R}\,\beta}\mathrm{arctanh}(g_3\,\beta)\,.
\end{equation}
Finally we can read off $\sigma$ from the $C$ equation in (\ref{eq:Cir0}):
\begin{equation}
e^{5 \sigma/2}\,=\, \dfrac{g_3}{16\,h}\,\dfrac{1}{\sqrt{1-g^2_3\,\beta^2}}\,.
\end{equation}

The cosmological constant reads
\begin{equation}\label{eq:cc-double}
\left( \dfrac{V_{\mu_\mathrm{L},\mu_\mathrm{R}}}{V_{0,0}} \right)^{5/4}\,=\, \dfrac{1}{1-g_3^2\beta^2}=\frac1{1-g_3^2\left(\dfrac{\kappa_\mathrm{L}^2}{g_\mathrm{L}^2}+\dfrac{\kappa_\mathrm{R}^2}{g_\mathrm{R}^2}\right)}
\end{equation}
generalizing (\ref{eq:cc-7d}). 

Along the lines of section \ref{sub:cc}, we can compare with the results in \cite{cremonesi-t}. (\ref{eq:crem}) has to be modified by adding the contribution from both $\mu_{L,R}$:
\begin{equation}\label{eq:crem-double}
	a_{\mu_\mathrm{L}, \mu_\mathrm{R}} = N^3\frac{k^2}{12} - N \frac k6 \left(\sum_a a^3 f_a^\mathrm{L}+ \sum_a a^3 f_a^\mathrm{R} \right)+ \ldots\,.
\end{equation}
The comparison with (\ref{eq:cc-double}) now works if one assumes 
\begin{equation}
	g_\mathrm{L}= g_\mathrm{R}\equiv g \, ,\qquad
	\frac{g_3^2}{g^2} = \frac1{N k^2}\,.
\end{equation}


\subsection{Two-tableau RG flows} 
\label{sub:double-flow}

To find domain walls connecting the vacua of the previous subsection, we proceed as we did in section \ref{sec:dw}: we modify the vacuum Ansatz by allowing all fields to depend on the radial coordinate $\rho$, and by no longer assuming that the $\phi^i$ are proportional to a reducible su(2) representation, only recovering this at $\rho\to \pm \infty$. 

Again we need to compute the quantities appearing in (\ref{eq:susy}). Since they now become a bit lengthy, we prefer writing them more compactly by defining
\begin{equation}
S_{RS}=
\left( 
\begin{array}{cc}
\delta_{rs}+\dfrac{\cosh \alpha-1}{\alpha^2}\,P_{rs}& \dfrac{\cosh \alpha-1}{\alpha^2}\,T_{r\,\hat s} \\[1.7ex]
\dfrac{\cosh \alpha-1}{\alpha^2}\,T_{\hat r\,s}& \delta_{\hat r\hat s}+\dfrac{\cosh \alpha-1}{\alpha^2}\,P_{\hat r\hat s}
\end{array}
\right)\,,
\end{equation}
which collects the four lower-right blocks in (\ref{eq:L-double}).

The only things we need to know about the pseudo-projector $S_{RS}$ is that it is invertible and that:
\begin{equation}
S_{RS}\,\phi^S_i\,=\, \cosh \alpha\,\phi_R^i \, ,\qquad \Rightarrow \qquad S^{-1}_{RS}\,\phi^S_i\,=\, \dfrac{1}{\cosh \alpha}\,\phi_R^i\,;
\end{equation}
recall that $\alpha^2\,=\,\alpha_\mathrm{L}^2\,+\,\alpha^2_\mathrm{R}$. We now have
\begin{equation}
\begin{split}
S^{-1}_{RS}\,P^{iS}_r\,&=\,-\dfrac{\sinh \alpha}{\alpha}\,\phi^i_R\,\dfrac{\partial_\rho\,\cosh \alpha}{\cosh \alpha}+\partial_\rho\left(\dfrac{\sinh \alpha}{\alpha}\,\phi^{i}_R\right)\,,\\
\sqrt{2}\,S^{-1}_{RS}\,C^{iS}\,&=\,-2\,g_3\,\dfrac{\sinh \alpha\,\cosh^2 \alpha}{\alpha}\,\phi^i_R\,+\,g_{L,R}\,\dfrac{\sinh^2 \alpha}{\alpha^2}\,\left[\phi^j, \phi^k\right]_R\,\epsilon^{ijk}\,,
\end{split}
\end{equation} 
where in the second line the choice $g_{L, R}$ depends whether the index $R$ is $r$ or $\hat r$. As in the single-tableau case, we find it useful to fix the radial gauge by taking $Q$ as in (\ref{eq:e-Q}). 

From $\delta \lambda^R=0$ we now obtain two copies of the Nahm-like equations (\ref{eq:nahm}):
\begin{equation}\label{eq:nahm-double}
\partial_\rho \Phi^i_{L, R}\,=\, -g_3\,\Phi^i_{L,R}\,+\,\dfrac{1}{2}\,\epsilon^{ijk}\,\left[ \Phi^i_{L,R}\,\Phi^j_{L,R} \right] \, ,\qquad
\Phi_{L,R}^i\equiv g_{L,R} \phi^i_{L,R}\dfrac{\tanh \alpha}{\alpha}\,.
\end{equation}
Using this, the analysis of $\delta \psi_\mu= 0=\delta\chi$ works just like in the single-tableau case, and we will not repeat it here.

Also, the analysis of (\ref{eq:nahm-double}) simply involves repeating the considerations of section \ref{sub:nahm} separately for the nilpotent elements $\mu_\mathrm{L}$, $\mu_\mathrm{R}$.



\section*{Acknowledgements}
We would like to thank F.~Apruzzi, B.~Assel, E.~Malek, N.~Mekareeya and A.~Zaffaroni for interesting discussions. 
The work of AG is supported by a Marie Sk\l{}odowska--Curie Individual Fellowship of the European Commission Horizon 2020 Program under contract number 702548 GaugedBH. GBDL, GLM and AT are supported in part by INFN.

\bibliography{at}
\bibliographystyle{at}

\end{document}